\documentclass{aa}  

\usepackage{graphicx}
\usepackage{txfonts}
\usepackage{lipsum}
\usepackage{lscape} 
\usepackage{float}
                                
\usepackage{placeins}           
                                
\usepackage{amsmath, amsfonts, amssymb}
\usepackage{longtable}
\usepackage{booktabs}
\usepackage{float}
\usepackage{subfigure}
\usepackage{color}
\usepackage{hyperref}
\hypersetup{colorlinks=true, linkcolor=black,citecolor=black,urlcolor=black}

\begin{document}

   \title{A Complete X-ray View of Supernova Remnant W28 with Einstein Probe: Spatial Distribution of Parameters and Origin of the Thermal-Composite Morphology}

   \subtitle{}

   \author{Yi-Heng Chi \inst{1} \and Ping Zhou\thanks{pingzhou@nju.edu.cn} \inst{1,2}
        \and Yang Chen\inst{1,2} \and Lei Sun\inst{1,2,3} \and Chengkui Li \inst{4} \and Shumei Jia \inst{4} \and Yong Chen \inst{4} \and Chong Ge \inst{5} \and Weimin Yuan \inst{6}
        }

   \institute{School of Astronomy and Space Science, Nanjing University, Nanjing 210023, China
             \and
             Key Laboratory of Modern Astronomy and Astrophysics, Nanjing University, Ministry of Education, Nanjing 210023, China
             \and Department of Astronomy, Tsinghua University, Beijing 100084, China
            \and State Key Laboratory of Particle Astrophysics, Institute of High Energy Physics, Chinese Academy of Sciences, Beijing 100049, China
            \and Department of Astronomy, Xiamen University, Xiamen, Fujian 361005, China
            \and National Astronomical Observatories, Chinese Academy of Sciences, Beijing 100012, China}

   \date{Received September 30, 20XX}
 
  \abstract{It has been an unsolved question what leads a supernova remnant (SNR) to a thermal composite rather than a typical shell-like morphology, and what causes recombining plasma inside it. With the 13-ks observation of the Following-up X-ray Telescope onboard the Einstein Probe, we give an overall X-ray picture of W28, one of the prototypical thermal composite SNRs. The observation revealed a shell-like structure west of W28 in radio, optical, and X-ray images, which may revise the known extent of the SNR to $72'\times45'$. Spectral analysis explicitly maps that the special relationship where the plasma experiences recombination in the interior of the remnant, spatially coincident with H$\alpha$ emissions, while in the other regions, the plasma is ionization-dominated. We found that W28 is generally isobaric from its center to the newly discovered shell, and it is even isothermal with a temperature of $\sim0.6$--0.7\,keV in the center before the cooling of the plasma. Saturated thermal conduction and cloud evaporation may cool down the plasma within $\sim3$\,kyr, the estimated recombination timescale. We revised the SNR dynamical age to $\sim8$\,kyr, much younger than previous estimates. The complex structure and complex ionization state distribution may suggest that centrally filled and shell-like morphologies coexist in W28. This state may depend on the environment in which the SNR evolves.}

\titlerunning{EP-FXT and XMM-Newton View of W28}
\authorrunning{Y.-H. Chi et al.}
   \maketitle\nolinenumbers 

\renewcommand{\arraystretch}{1.5}
\renewcommand{\dblfloatpagefraction}{.99}
\renewcommand{\textfraction}{.01}

\section{Introduction}
\label{sec: intro}
Supernova remnants (SNRs) are important extended objects mainly shaped by the interaction between supernova (SN) shocks and ambient gas, including the interstellar medium (ISM) in the environment and circumstellar medium (CSM) created by the stellar winds of the progenitor. The rapid shock of the SN can effectively compress, heat, and finally sweep the gas into a thin bright shell, which is the canonical morphology of SNRs and well expected by the self-similar models such as Sedov-Taylor solution \citep{Sedov1959, Taylor1950} and the Chevalier self-similar model \citep{Chevalier82}. The shock can also effectively accelerate particles to GeVs or even tens of TeVs, leading to synchrotron emissions from radio to X-rays on the shell. Apart from the shocked gas, neutron stars created in the core-collapse SNe can also prominently contribute to the total luminosity of the SNR with nonthermal emissions of their relativistic electron/positron plasma, or rather the pulsar wind nebula (PWN). The most representative example of PWN-dominated SNR (or filled center or plerion) is the Crab nebula. Also, composite SNRs are expected if both the shell and the PWN are observable. Such examples include G11.2$-$0.3, G292+1.8, and Kes~75.

Apart from the three morphologies mentioned above (shell-like, plerion, and composite), one new classification of SNR, thermal-composite (or mixed-morphology) SNRs, was discovered afterward, thanks to the development of X-ray astronomy. An archetype of this kind of SNRs is W28 (SNR~G6.4$-$0.1) at a distance of $\sim1.9$\,kpc \citep{Velazquez02}, characterized by centrally filled X-ray  and shell-like radio morphology \citep{Rho98}. Additionally, OH masers \citep{Frail94, Claussen97, Hoffman05}, molecular clouds \citep{Arikawa99, Velazquez02, Reach05, Mazumdar22}, and GeV-TeV $\gamma$-ray emissions probably with a hadronic origin \citep{Rowell00, Aharonian08, Abdo10, Li10, Cui18} are associated with W28, with a proposed age of $\sim33$--42\,kyr \citep{Rho02, Velazquez02, Li10}. X-ray observations towards the SNR center reveal that the electron temperature (measured from the bremsstrahlung continuum) is lower than the ionization temperature (reflected by the ionization state of heavy elements), meaning that the recombining process of ions dominates over ionization \citep{Sawada12}. This overionized or recombining non-equilibrium ionization (NEI) may suggest the plasma has experienced a cooling process from a high temperature \citep{Itoh89, Kawasaki05, Yamaguchi09, Pannuti17, Okon18, Himono23} or other mechanisms like photoionization \citep{Ono19}.

Some theoretical models are proposed to explain the unique morphology of thermal-composite SNRs. For very mature SNRs experiencing radiative loss, the cool shell would be invisible in X-rays due to the absorption \citep{Rho98}. Other scenarios generally require a high density environment with density gradients, including evaporation of cloudlets engulfed by the shock \citep{White91}, thermal conduction in the interior hot gas \citep{Cox99, Shelton99}, reflected shocks from a cavity wall \citep{Chen08}, and projection effect of an asymmetric shocked CSM shell \citep{Shimizu12}. For a comparison, the Sedov-Taylor solution assumes the evolution in a uniform medium.

However, a complex environment with density gradient may suggest different dynamical evolutions in different directions. In this case, a classification of SNRs could be expected as the transition between the shell-like and the thermal-composite, some part of which evolves in normal ISM while the other interacts with dense clouds. Meanwhile, the thermal-composite SNRs are commonly middle-aged, but their early-stage evolution is not clear yet. Especially for the nearby X-ray-bright SNRs with large angular sizes ($\gtrsim1^\circ$), a great proportion of them have thermal-composite X-ray morphology \citep{Ferrand12}, such as HB~3 \citep{Lazendic06}, HB~9 \citep{Sezer19}, HB~21 \citep{Lazendic06}, W63 \citep{Mavromatakis04}, and G65.3+5.7 \citep{Shelton04}. 

To understand the centrally filled X-ray morphology of thermal composite SNRs, one needs to map the properties of the shocked hot gas via spatially resolved X-ray analysis of the entire SNRs. However, current X-ray observation missions like Chandra \citep{Weisskopf02}, XMM-Newton \citep{Jasen01}, and Suzaku \citep{Mitsuda07} are limited to their small field of view ($17'$--$30'$). It is time-consuming to mosaic SNRs with a large angular size and multiple-pointing observations make it a great challenge to handle the spatially varying response and instrumental effect. Taking W28 for instance, previous studies \citep{Rho02, Kawasaki05, Sawada12, Zhou14, Nakamura14, Pannuti17, Okon18, Himono23} mainly focused on the W28 center, Clump D \citep[previously called southwestern shell;][]{Rho02}, and the NE shell (see the labels in Fig.~\ref{fig: xrayrgb}). Our new study covers the vicinity of these bright structures, which enables us to search for possible further extension of this remnant.

In this paper, we present the first spatially resolved spectral analysis of the entire W28 based on the Follow-up X-ray Telescope \citep[FXT,][]{Chen20, Chen25}, which is one of the two science instruments onboard the Einstein Probe \citep[EP, also named ``Tianguan'',][]{Yuan22} launched on January 9, 2024, in Xichang, China. The FXT consists of two co-aligned Wolter-I modules, FXTA \& FXTB. With a large field of $1^\circ\times1^\circ$, a total effective area of $\sim600$\,cm$^2$ at 1\,keV, and a low instrumental background in a level of $\sim4\times10^{-3}$\,counts\,s$^{-1}$\,cm$^{-2}$\,keV$^{-1}$  in 0.3--10\,keV \citep{Zhang22}, FXT has powerful capability of observing very diffuse X-ray sources such as SNRs, superbubbles, the Galactic ridge, and galaxy clusters. Despite its short exposure, FXT provides a brand new view of this famous SNR. Long-exposure XMM-Newton data are used as supplementary and verification in our study.

This article is organized as follows: we present the process of EP-FXT and XMM-Newton data in Sect.\,\ref{sec: data}. The morphology and spectroscopy analysis are shown in Sect.\,\ref{sec: mor} and Sect.\,\ref{sec: spec}. Then we discuss our results in Sect.\,\ref{sec: dis} and summarize this article in Sect.\,\ref{sec: sum}.

\section{Data reduction}\label{sec: data}
\subsection{EP-FXT}
W28 was proposed to verify the imaging spectroscopy ability of FXT in the performance verification (PV) phase from April to May 2024. The observational program consists of two individual pointing observations in the Full Frame mode with the Thin Filter (for detailed technical information of FXT, see the instrument webpage \footnote{http://epfxt.ihep.ac.cn/about} and the user guide \footnote{http://epfxt.ihep.ac.cn/downloads/FXT\_Users\_Guide\_v1\_20-3.pdf}), and the total coverage sky area is $1.5^\circ\times1^\circ$. The detailed information is listed in Appendix~\ref{app: obs}.

The FXT observations are reprocessed by the Follow-up X-ray Telescope Data Analysis Software (FXTDAS\footnote{http://epfxt.ihep.ac.cn/analysis}) with the standard reprocessing threads including scripts {\it fxtcoord}, {\it fxtpical}, {\it fxtparticleidentify}, {\it fxtcoord}, {\it fxtbadpix}, {\it fxthotpix}, {\it fxtgrade}, and {\it fxtgtigen}. In these processes, the ``bad events'' from bad/hot/flickering pixels, cosmic rays, and bad time intervals are removed. Energy- and coordinate-calibrated events are then re-constructed (for details see Chi et al. in prep.). Only events flagged with {\it status}=0 and {\it pattern}$\leq$12 are used for analysis. The event lists of two observations are reprojected and combined.

\begin{figure*}[ht]
    \centering
    \includegraphics[width=\textwidth]{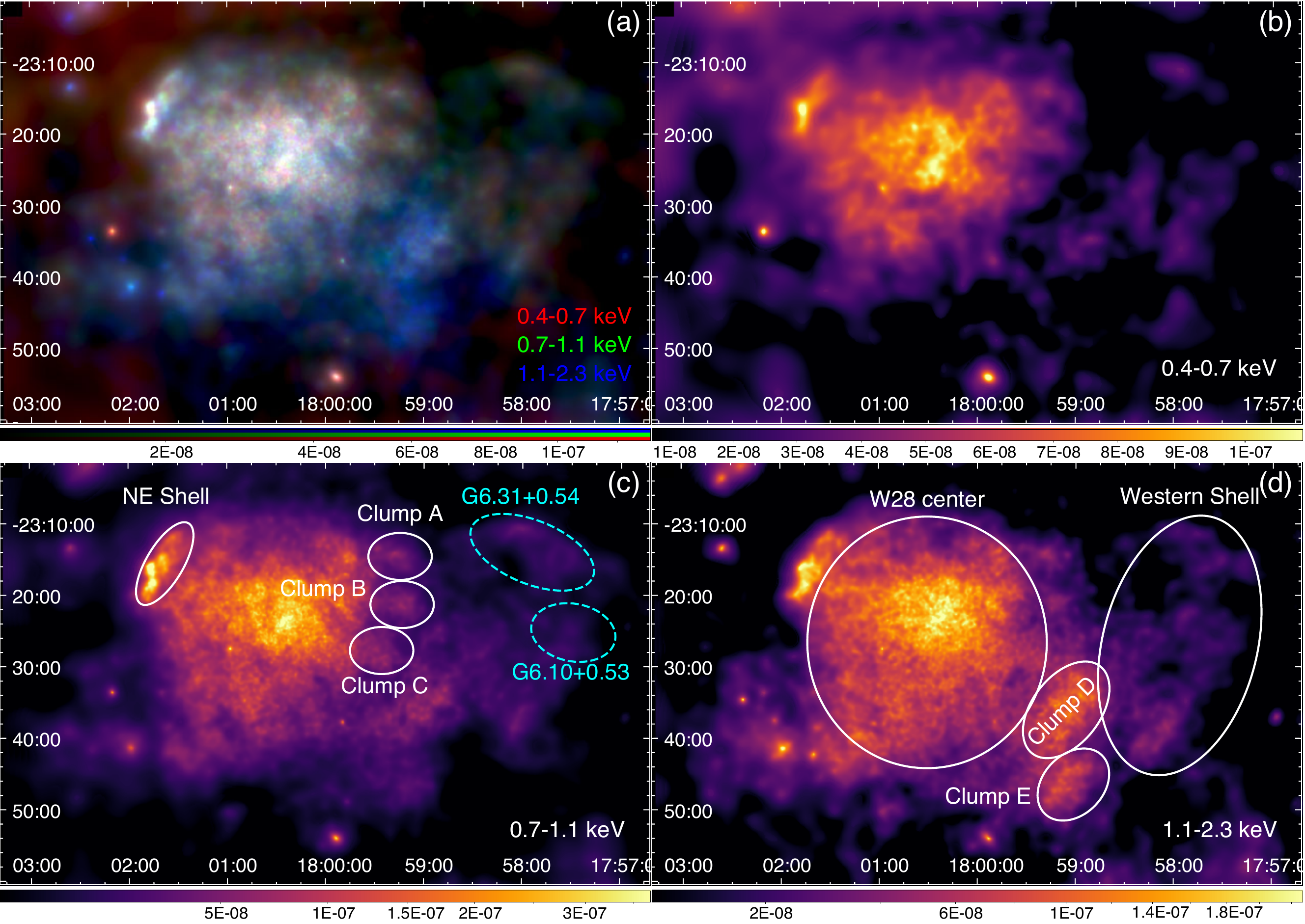}
    \caption{Energy-coded X-ray image of W28 complex with FXT (a), consisting of bands in 0.4--0.7\,keV (red), 0.7--1.1\,keV (green), and 1.1--2.3\,keV (blue). The single-band images of each energy band are shown in (b)(c)(d), respectively. The images are in units of count\,s$^{-1}$\,cm$^{-2}$ and in a square-root color scale to enhance weak emissions. All the images are vignetting-corrected and adaptively smoothed. Structures mentioned in this article are marked in (c) and (d). Cyan regions correspond to optical and/or radio SNR candidates \citep{Stupar11, Stupar18}.}
    \label{fig: xrayrgb}
\end{figure*}

Images of selected energy bands are extracted with the FTOOLS \citep{Blackburn95} script {\it ftcopy} from the reprojected event file. The task {\it fxtexpogen} creates exposure maps for vignetting correction and masking pixels with {\tt RAWY$<$}20 to avoid the ``shadow'' of the frame storage area. The exposure maps are combined regarding the effective area of FXTA and FXTB, and the vignetting-corrected image is adaptively smoothed with CIAO \citep[version 4.16,][]{Fruscione2006} script {\it dmimgadapt}. The vignetting-corrected and adaptively-smoothed images of W28 are shown in Fig.~\ref{fig: xrayrgb}.

Spectra of emission from the regions defined by polygons are extracted with Xselect and then grouped with {\it grppha} to guarantee a minimum signal-to-noise ratio of 3 per bin. The response files are generated with {\it fxtarfgen} and {\it fxtrmfgen}. Xspec \citep[version 12.14.a,][]{Arnaud96} based on AtomDB \citep[version 3.0.9,][]{Foster12} is used for the spectral fitting.

\subsection{XMM-Newton}
We also retrieve archival XMM-Newton data of W28. Compared to FXT, XMM-Newton has a slightly higher angular resolution and has higher throughout, especially in the W28 NE shell. The region of interest in our study consists of 13 different pointings with a total exposure of $\sim186$\,ks for an overall but not uniform coverage (see Appendix~\ref{app: obs} and Fig.~\ref{fig: xmmrgb}), most of which belong to large programs such as XMM-Newton Galactic Plane Survey \citep{Hands04}. This leads to difficulty in background estimation and response weighting for spectral analysis. Hence, we mainly used XMM-Newton for imaging here and used the NE shell of W28, which has the deepest XMM-Newton exposure, as a spectral calibrator to verify the performance of FXT (see Appendix~\ref{app: cal} for details).

The XMM-Newton data are processed and analyzed with XMM-Newton Science Analysis Software \citep[SAS version 20.0,][]{Gabriel04} and XMM-Newton Extended Source Analysis Software package (ESAS) with the latest Current Calibration Files. The MOS \citep{turner01} and pn \citep{Struder01} observations are calibrated with {\it emchain} and {\it epchain}, and then filtered the high-level background with {\it mos-filter} and {\it pn-filter}, respectively. We noticed that the XMM-Newton observation 0886110701, whose nominal point is in the southwest of W28, was coincidentally applied on the same day as the FXT observations. With a long exposure of $\sim20$\,ks, it can also play the role of a flare monitor for FXT, which shows a low smooth sky background count rate level.

Imaging and spectral extraction are based on threads {\it mos-spectra} and {\it pn-spectra}. The instrumental background (or quiescent particle background, QPB) images and spectra are modeled by {\it mos\_back} and {\it pn\_back}. The images are then mosaicked via {\it merge\_comp\_xmm} and then adaptively smoothed by {\it adapt\_merge}.

\begin{figure*}[ht]
    \centering
    \includegraphics[width=\textwidth]{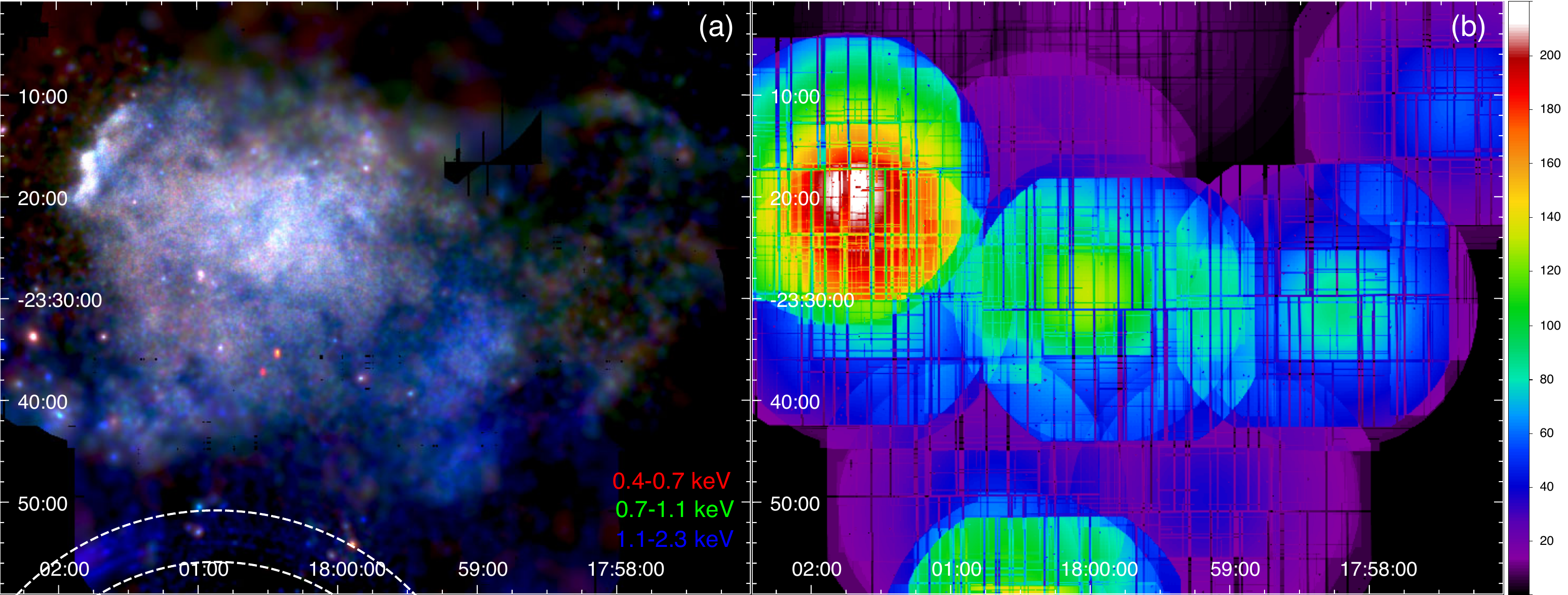}
    \caption{(a): The same as Fig.~\ref{fig: xrayrgb}(a) but based on the XMM-Newton data. The arc-like patterns of the stray light from GX~5$-$1 is visible in the south, marked between white dashed lines. (b): The effective exposure time of XMM-Newton in units of ks scaled as MOS2 with the Medium filter, which share similar effective area with one FXT module.}
    \label{fig: xmmrgb}
\end{figure*}

\begin{figure*}[ht]
    \centering
    \includegraphics[width=\textwidth]{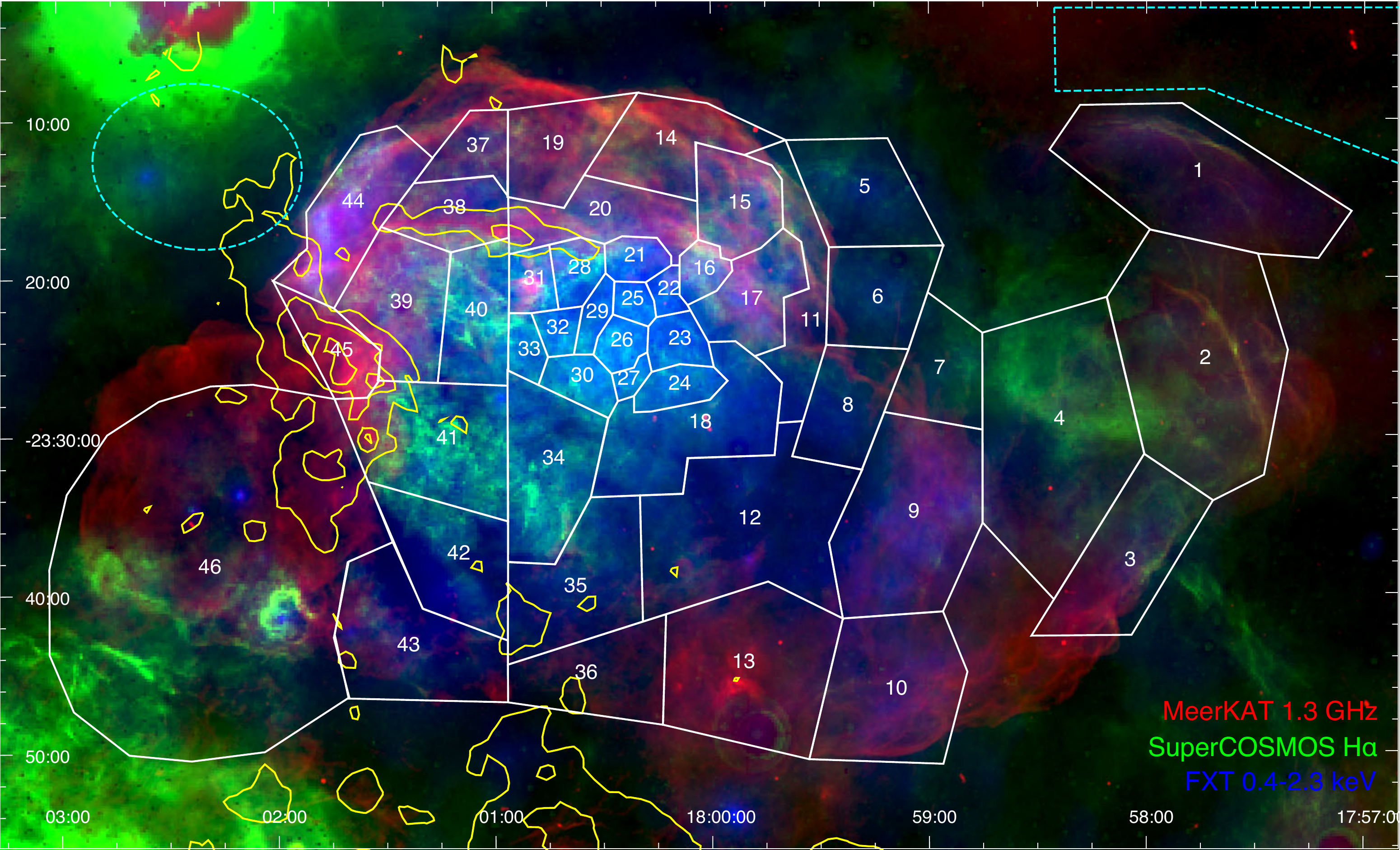}
    \caption{Tri-color image of W28 in radio (red, 1.3\,GHz continuum), H$\alpha$ (green), and X-ray (blue, 0.4--2.3\,keV) bands. A square root color scale is used to highlight faint diffuse emissions in radio and X-ray images. White polygon regions are used for spectral extraction and analysis and cyan dashed regions correspond to the sky background. The yellow contours correspond to the MWISP CO intensity in the $-20$--10\,km\,s$^{-1}$ \citep{Tu24}. Optical nebulae southeast and northeast of W28 are the Lagoon Nebula Messier~8 and the Trifid Nebula Messier~20, respectively.}
    \label{fig: 3bandrgb}
\end{figure*}

\subsection{Optical and radio data}
Optical and radio data are used to reveal the multi-wavelength morphology of W28. We retrieved narrowband H$\alpha$ images of the AAO/UKST SuperCOSMOS H$\alpha$ survey with an angular resolution of $\sim1''$ \citep{Parker05}. To underline the morphology of H$\alpha$-emitting gas from the starlight continuum, the R band images are used to make an H$\alpha$/R ratio map. In radio, the SARAO MeerKAT 1.3\,GHz Galactic Plane Survey \citep{Goedhart24} that covers areas of $|b|\leq1.5^\circ$ provides complete coverage of W28 along with its environment with an angular resolution of $\sim8''$.

\section{Results}
\subsection{Morphology}
\label{sec: mor}
Fig.~\ref{fig: xrayrgb} and \ref{fig: xmmrgb} show the X-ray morphology of W28. The centrally filled plasma with a radius of $\sim15'$ (or W28 center hereafter, defined in Fig.~\ref{fig: xrayrgb}) dominates X-ray emissions of the entire SNR, which is the canonical feature of thermal-composite SNRs. The X-ray luminosity peaks in the center and there are no evident asymmetric structures like jets or bars. Two clumpy structures \citep[Clumps D \& E; southwestern shell in][]{Rho02} are enhanced above 1.1\,keV in the southwest. The northern one, Clump D, looks brighter in the hard X-ray band. Either a strong foreground absorption or a high temperature may account for this feature. 

Previous studies \citep{Long91, Zhou14, Nakamura14, Pannuti17, Okon18, Himono23} focused on the center, the northeastern shell, and the brighter clump in the southwest.
However, FXT reveals a much more extended X-ray morphology than before. Two shell-like X-ray structures (collectively named ``western shell'' hereafter) are evident to the west of the W28 center, whose extension is comparable to the diameter of the center. This structure is spatially coincident with radio and optical filaments in the multi-wavelength image (Fig.~\ref{fig: 3bandrgb}). Such a morphology is typically formed by a shock interaction with the ISM and is common in SNRs. Noticeably, the northern part of this western shell was reported as part of an undiscovered SNR (G6.31+0.54 therein), diagnosed by an enhanced [S\,{\footnotesize II}]/H$\alpha$ ratio \citep{Stupar18}. Meanwhile, its central part was proposed as a Galactic SNR G6.10+0.53 according to H$\alpha$ image and radio spectral index \citep{Brogan06, Stupar11}. Here, the FXT image provides further evidence for its SNR origin through the spectral analysis (see the subsection below). Without a complete shell, the western shell could be part of W28 according to its curvature and orientation. Similar morphology appears in the famous thermal-composite SNR IC 443, whose western shell, located farther away, is fainter and larger than its eastern shell.

Besides, three X-ray-emitting clumps (Clump A, B, and C, or Region 5, 6, and 8 in Fig.~\ref{fig: 3bandrgb}) appear to the west of the center aligned from north to south, and share a similar size, $\sim6'$ in diameter, seemingly outside the bright radio shell. Clumps A and B are located out of the radio shell and the H$\alpha$ emissions there do not show filamentary structures in shocked regions like the NE shell and the western shell.

In the east, diffuse X-ray emissions (region 46) are visible near a radio ring, which was considered an independent SNR, G6.5$-$0.4. However, the X-rays appear south of the radio emissions. Therefore, the morphology alone cannot tell whether the X-ray emissions here originate from G6.5$-$0.4 or W28. A point-like source (R.A.=18:02:00, Decl.=$-$23:42:00 in Fig.~\ref{fig: xrayrgb}(d)) with a diffuse X-ray ``halo'' around it appears south of the eastern X-ray emissions. These features are spatially associated with an O6III blue giant in an HII region according to the SIMBAD database \citep{Wenger00}. Related optical and radio emissions can be seen in Fig.~\ref{fig: 3bandrgb} as well.

\subsection{Spectroscopy}
\label{sec: spec}
We manually divided emissions associated (or overlapped) with W28 into 46 polygon regions (see Fig.~\ref{fig: 3bandrgb}) for spectrum extraction. The definition of a region mainly depends on its morphology to guarantee a uniform distribution of its X-ray surface brightness and the total photon numbers (smaller area for brighter region). Besides, the region boundaries are limited within the field of view in order to make full use of the observed X-ray photons. The spatial sampling rate declines from the center to the edge to balance the spectral quality and spatial resolution of the spectral analysis. We directly subtract the sky background (dashed cyan ellipse in Fig.~\ref{fig: 3bandrgb}) from the source spectra, as W28 is bright enough (1--2 orders of magnitude higher than the sky in 1--2\,keV) and the sky background is difficult to model for the short exposure. 

\begin{figure}[ht]
    \centering
    \includegraphics[width=\linewidth]{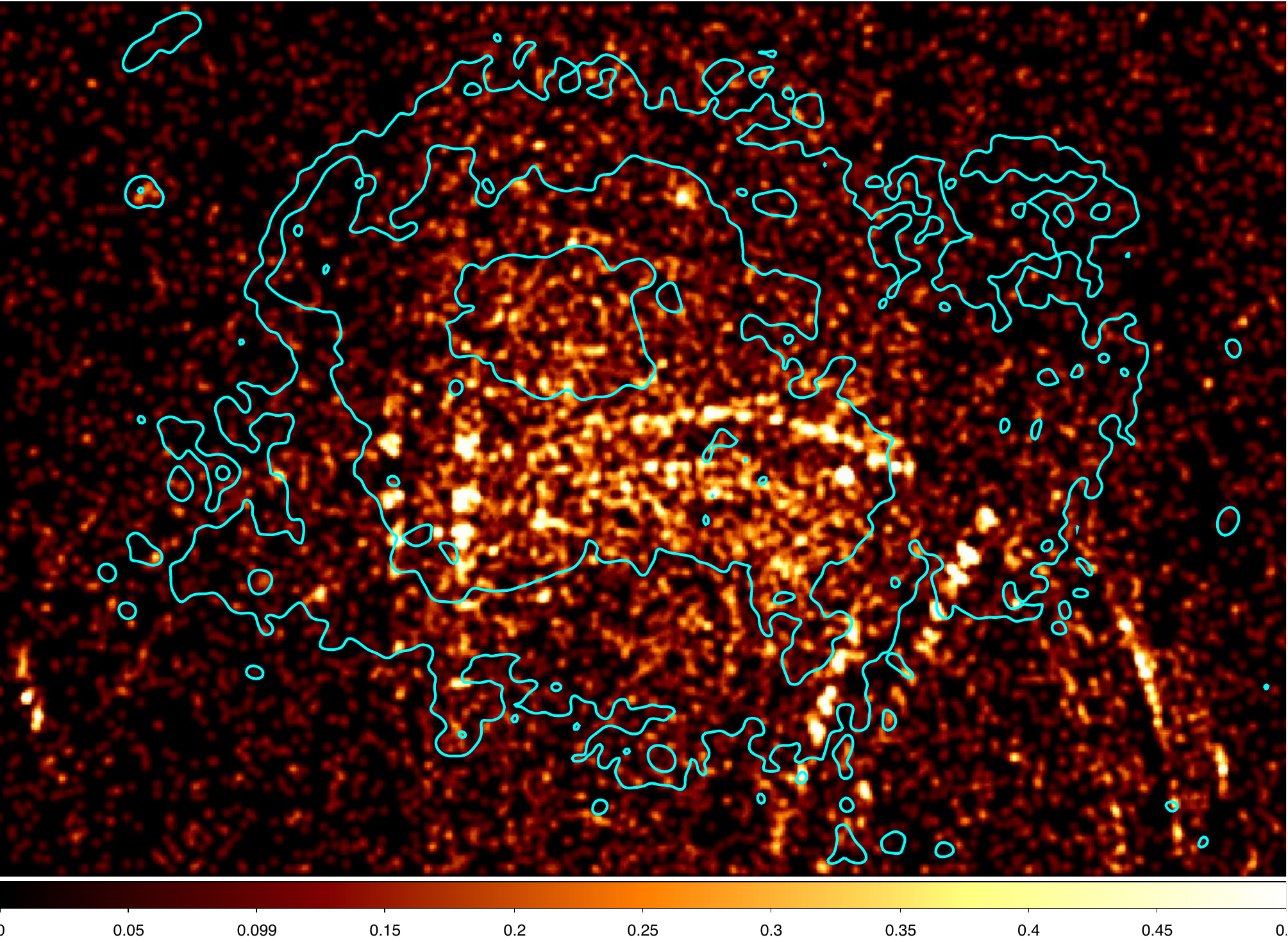}
    \caption{Counts map (not vignetting-corrected) in 3--7\,keV to show the effect of the stray light. The cyan contour corresponds to 0.5, 2, and 8 counts for W28 in 0.4--2\,keV.}
    \label{fig: stray}
\end{figure}

The challenge for spectral analysis is the contamination of the stray light, or rather the X-ray photons experiencing only one reflection from an off-axis bright source or even several degrees out of the field of view. The stray light, consisting of filamentary or knot-like patterns, in the lower part of the field of view is from the bright X-ray binary GX~5$-$1 to the south of W28, especially dominant in hard X-ray bands above 3\,keV (Fig.~\ref{fig: stray}). Besides, we extract the spectra from a region to the southwest of the remnant to evaluate the stray light spectra, which were found quite flat and reproduced with a double-broken {\it powerlaw} model ({\it bkn2pow}). When fitting the SNR spectra in different regions, we fix the shape of the stray light component but free its normalization as a component to explain the excess above $\sim2$\,keV. Fortunately, as we can see in Fig.~\ref{fig: xrayrgb}, lower-energy X-ray photons are still dominated by the plasma of W28, but admittedly, stray light may mislead the fitting result if there is a hard component from hot plasma or nonthermal emissions (see details at the end of this section).

We only use FXT data to perform spectral analysis across the SNR. We consider several models in the spectral fitting. First, the overionized NEI model {\it vrnei} has found to characterize the X-ray-emitting plasma in W28 center \citep{Zhou14, Pannuti17, Okon18, Himono23}. The shocked underionized plasma is described with the {\it vnei} model, which along with {vrnei} assumes a constant temperature and single ionization/recombination parameter. Finally, the {\it vapec} model is used for plasma in CIE. 

All the models have variable metallicities, of which the Ne, Mg, and Si abundances can always be well constrained with good statistics. O, Fe, and S abundances can only be measured in part of the spectra with evident emission features. We first applied the solar abundance \citep{Wilms00} by default but found that in most cases the metals tend to be subsolar. This is consistent with the majority of previous studies of W28 \citep{Zhou14, Nakamura14, Himono23} although oversolar abundance was also reported \citep{Okon18}. Therefore we finally use 0.3\,Z$_\odot$ as the default metal abundance of C, N, O, Ne, Mg, Si, S, Ar, Ca, Fe, and Ni, the result of solar abundance settings is also listed in the Appendix for reference. The distribution of most parameters is generally similar between the two model settings. We find that for almost all the regions, one thermal component is enough for providing a good fit with the reduced chi-square $\chi^2_r\lesssim1.2$.

\longtab{
\setlength\tabcolsep{3pt}
{
\small
\begin{longtable}{ccccccccccccc}
    \caption{Fitting results (Z=0.3\,Z$_\odot$) with $1\sigma$ error.}
    \label{tab: fit3} \\
    \hline
        Region & $N_\mathrm{H}$ & $kT$ & $kT_\mathrm{init}$ & O & Ne & Mg & Si & S & Fe & $\tau^a$ & $n_\mathrm{e}n_\mathrm{H}l$ & $\chi^2_r$/dof \\
        ~ & ($10^{21}$\,cm$^{-2}$) & (keV) & (keV) & (Z/Z$_\odot$) & (Z/Z$_\odot$) & (Z/Z$_\odot$) & (Z/Z$_\odot$) & (Z/Z$_\odot$) & (Z/Z$_\odot$) & ($10^{10}$\,s\,cm$^{-3}$) & (cm$^{-6}$\,pc) & ~ \\ 
        \hline
    \endfirsthead
    \caption{(continued)} \\ \hline
        Region & $N_\mathrm{H}$ & $kT$ & $kT_\mathrm{init}$ & O & Ne & Mg & Si & S & Fe & $\tau^a$ & $n_\mathrm{e}n_\mathrm{H}l$ & $\chi^2_r$/dof \\
        ~ & ($10^{21}$\,cm$^{-2}$) & (keV) & (keV) & (Z/Z$_\odot$) & (Z/Z$_\odot$) & (Z/Z$_\odot$) & (Z/Z$_\odot$) & (Z/Z$_\odot$) & (Z/Z$_\odot$) & ($10^{10}$\,s\,cm$^{-3}$) & (cm$^{-6}$\,pc) & ~ \\ 
        \hline
    \endhead
    \hline
    \endfoot
    \hline
        \multicolumn{13}{l}{{\bf Notes:} Some regions can be fitted with more than one model sets. Their results are maintained here for reference. } \\
        \multicolumn{13}{l}{$^a$Ionization timescale $\tau=n_\mathrm{e}t$. No number means CIE. } \\
        \multicolumn{13}{l}{$^b$ Not shown in Fig.~\ref{fig: resultmap3}.} \\
    \endlastfoot
1$^{b}$ & $7.6_{-1.7}^{+1.7}$ & $0.47_{-0.17}^{+0.23}$ & $\cdots$ & $\cdots$ & $0.30_{-0.17}^{+0.12}$ & $0.22_{-0.11}^{+0.13}$ & $0.90_{-0.32}^{+0.89}$ & $\cdots$ & $0.09_{-0.05}^{+0.05}$ & $>5.42$ & $0.46_{-0.14}^{+1.17}$ & 1.04/54 \\
1 & $7.6_{-1.7}^{+4.2}$ & $1.25_{-0.89}^{+0.31}$ & $\cdots$ & $\cdots$ & $0.45_{-0.22}^{+0.16}$ & $0.40_{-0.25}^{+0.17}$ & $0.87_{-0.34}^{+2.13}$ & $\cdots$ & $\cdots$ & $1.02_{-0.24}^{+0.63}$ & $0.10_{-0.05}^{+1.74}$ & 1.06/55 \\
2 & $1.77_{-0.65}^{+0.79}$ & $1.67_{-0.41}^{+1.13}$ & $\cdots$ & $<0.03$ & $0.09_{-0.07}^{+0.06}$ & $0.24_{-0.09}^{+0.11}$ & $0.74_{-0.17}^{+0.20}$ & $\cdots$ & $0.08_{-0.03}^{+0.04}$ & $2.9_{-1.2}^{+2.2}$ & $0.078_{-0.022}^{+0.025}$ & 1.07/96 \\
3 & $8.0_{-1.0}^{+1.1}$ & $1.87_{-0.70}^{+1.81}$ & $\cdots$ & $\cdots$ & $0.31_{-0.13}^{+0.10}$ & $0.22_{-0.10}^{+0.14}$ & $0.28_{-0.12}^{+0.14}$ & $\cdots$ & $0.16_{-0.08}^{+0.11}$ & $1.99_{-0.67}^{+1.39}$ & $0.14_{-0.05}^{+0.09}$ & 0.93/111 \\
4 & $8.19_{-0.29}^{+0.51}$ & $1.36_{-0.24}^{+0.14}$ & $\cdots$ & $\cdots$ & $0.26_{-0.03}^{+0.02}$ & $0.28_{-0.05}^{+0.05}$ & $0.68_{-0.16}^{+0.19}$ & $\cdots$ & $0.44_{-0.14}^{+0.20}$ & $0.59_{-0.08}^{+0.13}$ & $0.28_{-0.03}^{+0.09}$ & 1.12/238 \\
5 & $7.4_{-1.5}^{+1.7}$ & $0.86_{-0.32}^{+0.54}$ & $\cdots$ & $\cdots$ & $\cdots$ & $0.60_{-0.13}^{+0.15}$ & $1.05_{-0.34}^{+0.51}$ & $\cdots$ & $\cdots$ & $1.6_{-0.6}^{+1.9}$ & $0.24_{-0.11}^{+0.30}$ & 0.89/58 \\
6 & $4.8_{-1.2}^{+1.3}$ & $0.68_{-0.06}^{+0.07}$ & $\cdots$ & $\cdots$ & $0.19_{-0.08}^{+0.09}$ & $0.59_{-0.15}^{+0.20}$ & $0.61_{-0.20}^{+0.25}$ & $\cdots$ & $0.25_{-0.07}^{+0.09}$ & $14.92_{-5.32}^{+9.53}$ & $0.31_{-0.07}^{+0.09}$ & 1.02/106 \\
7 & $6.8_{-1.7}^{+1.9}$ & $0.70_{-0.06}^{+0.08}$ & $\cdots$ & $\cdots$ & $<0.42$ & $0.53_{-0.19}^{+0.26}$ & $0.85_{-0.20}^{+0.26}$ & $\cdots$ & $0.13_{-0.06}^{+0.12}$ & $\cdots$ & $0.50_{-0.10}^{+0.14}$ & 0.76/74 \\
8 & $5.9_{-1.3}^{+1.3}$ & $0.70_{-0.05}^{+0.06}$ & $\cdots$ & $0.87_{-0.77}^{+2.53}$ & $0.37_{-0.19}^{+0.50}$ & $0.44_{-0.14}^{+0.31}$ & $0.61_{-0.08}^{+0.30}$ & $\cdots$ & $0.08_{-0.04}^{+0.11}$ & $\cdots$ & $0.78_{-0.24}^{+0.17}$ & 1.27/172 \\
9 & $15.19_{-0.49}^{+0.54}$ & $0.19_{-0.01}^{+0.01}$ & $1.66_{-0.25}^{+0.41}$ & $0.59_{-0.12}^{+0.17}$ & $0.08_{-0.03}^{+0.04}$ & $0.35_{-0.05}^{+0.08}$ & $0.24_{-0.04}^{+0.05}$ & $\cdots$ & $0.62_{-0.24}^{+0.58}$ & $46.35_{-4.45}^{+5.07}$ & $35.4_{-7.2}^{+8.2}$ & 1.18/466 \\
10 & $18.45_{-0.60}^{+1.19}$ & $0.80_{-0.07}^{+0.14}$ & $\cdots$ & $\cdots$ & $0.31_{-0.15}^{+0.21}$ & $0.37_{-0.08}^{+0.08}$ & $0.47_{-0.08}^{+0.08}$ & $\cdots$ & $\cdots$ & $18.9_{-9.7}^{+11.4}$ & $1.82_{-0.41}^{+0.35}$ & 1.03/142 \\
11 & $4.27_{-0.67}^{+0.78}$ & $0.67_{-0.05}^{+0.04}$ & $\cdots$ & $\cdots$ & $0.22_{-0.10}^{+0.20}$ & $0.55_{-0.13}^{+0.19}$ & $0.53_{-0.12}^{+0.14}$ & $\cdots$ & $0.14_{-0.03}^{+0.04}$ & $44_{-20}^{+59}$ & $0.58_{-0.09}^{+0.12}$ & 1.05/166 \\
12 & $10.20_{-0.69}^{+0.65}$ & $0.21_{-0.02}^{+0.02}$ & $1.78_{-0.44}^{+0.84}$ & $\cdots$ & $0.12_{-0.03}^{+0.05}$ & $0.27_{-0.04}^{+0.05}$ & $0.23_{-0.04}^{+0.07}$ & $0.14_{-0.06}^{+0.10}$ & $0.09_{-0.05}^{+0.05}$ & $30.75_{-5.24}^{+4.47}$ & $11.2_{-2.4}^{+2.1}$ & 1.00/421 \\
13 & $14.6_{-1.9}^{+2.1}$ & $1.35_{-0.31}^{+0.48}$ & $\cdots$ & $\cdots$ & $0.27_{-0.20}^{+0.28}$ & $0.36_{-0.11}^{+0.14}$ & $0.32_{-0.08}^{+0.09}$ & $\cdots$ & $\cdots$ & $5.25_{-1.88}^{+4.83}$ & $0.40_{-0.12}^{+0.18}$ & 1.01/127 \\
14 & $3.2_{-1.8}^{+1.8}$ & $1.36_{-0.53}^{+1.85}$ & $\cdots$ & $\cdots$ & $0.55_{-0.18}^{+0.20}$ & $0.91_{-0.25}^{+0.35}$ & $0.71_{-0.42}^{+0.58}$ & $\cdots$ & $\cdots$ & $3.19_{-1.45}^{+3.57}$ & $0.09_{-0.04}^{+0.07}$ & 1.01/76 \\
15 & $7.1_{-1.0}^{+1.1}$ & $0.35_{-0.03}^{+0.03}$ & $\cdots$ & $0.28_{-0.16}^{+0.43}$ & $0.24_{-0.07}^{+0.14}$ & $0.30_{-0.07}^{+0.12}$ & $0.59_{-0.14}^{+0.25}$ & $\cdots$ & $0.12_{-0.04}^{+0.08}$ & $\cdots$ & $4.84_{-1.19}^{+1.64}$ & 1.00/218 \\
16 & $4.57_{-0.53}^{+0.58}$ & $0.69_{-0.03}^{+0.03}$ & $\cdots$ & $\cdots$ & $0.45_{-0.14}^{+0.15}$ & $0.39_{-0.12}^{+0.13}$ & $0.33_{-0.10}^{+0.11}$ & $\cdots$ & $0.17_{-0.03}^{+0.04}$ & $\cdots$ & $2.71_{-0.34}^{+0.40}$ & 1.08/157 \\
17 & $4.97_{-0.53}^{+0.53}$ & $0.63_{-0.03}^{+0.03}$ & $\cdots$ & $0.33_{-0.22}^{+0.33}$ & $0.29_{-0.08}^{+0.10}$ & $0.43_{-0.08}^{+0.10}$ & $0.28_{-0.06}^{+0.06}$ & $\cdots$ & $0.11_{-0.03}^{+0.04}$ & $\cdots$ & $2.10_{-0.25}^{+0.27}$ & 1.04/345 \\
18 & $8.42_{-0.40}^{+0.34}$ & $0.16_{-0.01}^{+0.01}$ & $0.75_{-0.03}^{+0.04}$ & $\cdots$ & $0.08_{-0.01}^{+0.01}$ & $0.22_{-0.03}^{+0.04}$ & $0.23_{-0.03}^{+0.04}$ & $0.36_{-0.17}^{+0.20}$ & $0.31_{-0.09}^{+0.15}$ & $8.63_{-3.25}^{+3.74}$ & $26.1_{-3.4}^{+3.4}$ & 1.14/548 \\
19 & $5.0_{-1.5}^{+1.6}$ & $0.65_{-0.05}^{+0.05}$ & $\cdots$ & $\cdots$ & $0.47_{-0.22}^{+0.36}$ & $0.77_{-0.27}^{+0.44}$ & $0.55_{-0.24}^{+0.34}$ & $\cdots$ & $\cdots$ & $\cdots$ & $0.59_{-0.13}^{+0.17}$ & 1.05/81 \\
20 & $4.26_{-6.3}^{+5.7}$ & $0.70_{-0.01}^{+0.03}$ & $\cdots$ & $0.75_{-0.45}^{+0.64}$ & $0.15_{-0.06}^{+0.09}$ & $0.42_{-0.10}^{+0.12}$ & $0.36_{-0.08}^{+0.11}$ & $\cdots$ & $0.18_{-0.04}^{+0.06}$ & $29.6_{-9.3}^{+13.1}$ & $1.20_{-0.16}^{+0.15}$ & 1.01/276 \\
21 & $7.14_{-0.40}^{+0.39}$ & $0.15_{-0.01}^{+0.01}$ & $0.62_{-0.05}^{+0.05}$ & $0.15_{-0.07}^{+0.07}$ & $0.05_{-0.02}^{+0.02}$ & $0.18_{-0.04}^{+0.06}$ & $0.12_{-0.06}^{+0.08}$ & $\cdots$ & $\cdots$ & $10.4_{-2.1}^{+2.1}$ & $88.6_{-25.8}^{+32.1}$ & 1.10/199 \\
21$^{b}$ & $3.13_{-0.50}^{+0.13}$ & $0.66_{-0.05}^{+0.06}$ & $\cdots$ & $0.21_{-0.19}^{+0.30}$ & $0.18_{-0.07}^{+0.13}$ & $0.31_{-0.08}^{+0.10}$ & $0.27_{-0.08}^{+0.09}$ & $\cdots$ & $0.09_{-0.02}^{+0.03}$ & $36.1_{-14.7}^{+34.6}$ & $3.27_{-0.50}^{+0.68}$ & 1.13/199 \\
22 & $4.15_{-0.84}^{+1.19}$ & $0.65_{-0.06}^{+0.06}$ & $\cdots$ & $\cdots$ & $0.24_{-0.06}^{+0.08}$ & $0.49_{-0.11}^{+0.14}$ & $0.57_{-0.15}^{+0.18}$ & $\cdots$ & $0.16_{-0.03}^{+0.04}$ & $24.7_{-7.3}^{+13.2}$ & $2.83_{-0.55}^{+0.88}$ & 1.06/119 \\
23 & $8.10_{-1.27}^{+0.63}$ & $0.15_{-0.02}^{+0.05}$ & $0.72_{-0.06}^{+0.11}$ & $0.53_{-0.18}^{+0.69}$ & $0.10_{-0.04}^{+0.18}$ & $0.29_{-0.10}^{+0.49}$ & $0.31_{-0.12}^{+0.46}$ & $\cdots$ & $0.57_{-0.31}^{+0.75}$ & $7.5_{-6.3}^{+8.1}$ & $34.5_{-23.1}^{+21.3}$ & 1.12/258 \\
24 & $7.50_{-0.40}^{+0.35}$ & $0.16_{-0.01}^{+0.01}$ & $0.83_{-0.07}^{+0.11}$ & $0.38_{-0.08}^{+0.11}$ & $0.04_{-0.02}^{+0.03}$ & $0.24_{-0.05}^{+0.09}$ & $0.14_{-0.05}^{+0.08}$ & $\cdots$ & $\cdots$ & $9.47_{-2.31}^{+2.18}$ & $36.0_{-10.3}^{+11.8}$ & 1.09/238 \\
24$^{b}$ & $7.62_{-0.58}^{+0.49}$ & $0.15_{-0.01}^{+0.02}$ & $0.79_{-0.06}^{+0.09}$ & $\cdots$ & $0.04_{-0.02}^{+0.02}$ & $0.20_{-0.04}^{+0.05}$ & $0.11_{-0.04}^{+0.05}$ & $\cdots$ & $0.31_{-0.13}^{+0.30}$ & $10.5_{-5.0}^{+6.9}$ & $46.6_{-8.5}^{+8.7}$ & 1.09/238 \\
25$^{b}$ & $5.01_{-1.01}^{+0.85}$ & $0.43_{-0.09}^{+0.13}$ & $1.01_{-0.15}^{+1.18}$ & $\cdots$ & $0.51_{-0.22}^{+0.24}$ & $1.18_{-0.41}^{+0.43}$ & $0.69_{-0.25}^{+0.43}$ & $\cdots$ & $0.16_{-0.03}^{+0.05}$ & $9.1_{-4.8}^{+29.5}$ & $7.18_{-3.27}^{+5.93}$ & 1.05/168 \\
25 & $6.71_{-0.73}^{+0.56}$ & $0.19_{-0.03}^{+0.07}$ & $0.76_{-0.06}^{+1.03}$ & $0.50_{-0.20}^{+0.20}$ & $0.20_{-0.09}^{+1.04}$ & $0.49_{-0.21}^{+1.67}$ & $0.38_{-0.17}^{+0.38}$ & $\cdots$ & $\cdots$ & $5.5_{-2.5}^{+2.5}$ & $27.2_{-15.6}^{+20.2}$ & 1.07/168 \\
26 & $7.60_{-0.48}^{+0.42}$ & $0.15_{-0.01}^{+0.01}$ & $0.73_{-0.04}^{+0.05}$ & $\cdots$ & $0.04_{-0.02}^{+0.02}$ & $0.17_{-0.03}^{+0.04}$ & $0.21_{-0.05}^{+0.06}$ & $\cdots$ & $0.43_{-0.18}^{+0.36}$ & $10.5_{-4.7}^{+5.5}$ & $82.9_{-13.4}^{+13.7}$ & 1.05/260 \\
27 & $5.23_{-0.81}^{+0.79}$ & $0.23_{-0.07}^{+0.08}$ & $1.12_{-0.18}^{+0.19}$ & $\cdots$ & $<0.17$ & $0.30_{-0.16}^{+0.31}$ & $<0.29$ & $\cdots$ & $0.14_{-0.05}^{+0.08}$ & $5.7_{-3.2}^{+3.4}$ & $23.0_{-10.1}^{+17.9}$ & 0.98/95 \\
28 & $6.48_{-0.68}^{+0.69}$ & $0.20_{-0.04}^{+0.07}$ & $0.66_{-0.05}^{+0.07}$ & $0.28_{-0.11}^{+0.28}$ & $0.08_{-0.04}^{+0.10}$ & $0.25_{-0.10}^{+0.20}$ & $0.38_{-0.14}^{+0.23}$ & $\cdots$ & $0.16_{-0.08}^{+0.19}$ & $<7.16$ & $35.4_{-17.0}^{+24.2}$ & 1.15/223 \\
29 & $5.77_{-0.86}^{+0.90}$ & $0.51_{-0.07}^{+0.08}$ & $1.20_{-0.24}^{+0.90}$ & $0.41_{-0.21}^{+0.88}$ & $0.81_{-0.35}^{+0.70}$ & $1.00_{-0.35}^{+0.64}$ & $0.99_{-0.28}^{+0.48}$ & $\cdots$ & $0.26_{-0.09}^{+0.20}$ & $24.1_{-10.5}^{+18.2}$ & $4.0_{-1.3}^{+2.3}$ & 1.19/161 \\
30 & $7.26_{-0.52}^{+0.47}$ & $0.19_{-0.02}^{+0.02}$ & $0.69_{-0.05}^{+0.07}$ & $0.41_{-0.13}^{+0.25}$ & $0.10_{-0.04}^{+0.08}$ & $0.31_{-0.09}^{+0.15}$ & $0.26_{-0.09}^{+0.16}$ & $\cdots$ & $0.31_{-0.13}^{+0.26}$ & $5.9_{-2.9}^{+3.9}$ & $34.5_{-12.7}^{+15.0}$ & 1.02/263 \\
31 & $7.02_{-0.42}^{+0.43}$ & $0.19_{-0.02}^{+0.02}$ & $0.84_{-0.12}^{+0.26}$ & $0.37_{-0.11}^{+0.11}$ & $0.14_{-0.06}^{+0.07}$ & $0.23_{-0.07}^{+0.10}$ & $0.26_{-0.09}^{+0.13}$ & $\cdots$ & $\cdots$ & $10.2_{-2.8}^{+3.4}$ & $33.3_{-9.1}^{+16.7}$ & 1.05/207 \\
32 & $8.54_{-0.68}^{+0.66}$ & $0.19_{-0.02}^{+0.03}$ & $0.68_{-0.07}^{+0.09}$ & $0.38_{-0.16}^{+0.44}$ & $<0.09$ & $0.24_{-0.08}^{+0.19}$ & $0.40_{-0.15}^{+0.29}$ & $\cdots$ & $0.40_{-0.21}^{+0.59}$ & $8.5_{-4.2}^{+6.9}$ & $51.8_{-26.5}^{+33.8}$ & 1.07/164 \\
33 & $4.55_{-0.79}^{+1.23}$ & $0.52_{-0.17}^{+0.07}$ & $>0.66$ & $\cdots$ & $0.17_{-0.16}^{+0.16}$ & $0.73_{-0.23}^{+0.28}$ & $1.03_{-0.70}^{+0.35}$ & $\cdots$ & $0.21_{-0.05}^{+0.09}$ & $31.2_{-21.7}^{+11.6}$ & $3.88_{-0.96}^{+0.98}$ & 1.12/155 \\
34 & $7.42_{-0.37}^{+0.40}$ & $0.17_{-0.01}^{+0.01}$ & $0.74_{-0.04}^{+0.05}$ & $0.34_{-0.07}^{+0.10}$ & $0.09_{-0.02}^{+0.03}$ & $0.36_{-0.07}^{+0.08}$ & $0.29_{-0.06}^{+0.08}$ & $0.52_{-0.26}^{+0.32}$ & $0.30_{-0.12}^{+0.15}$ & $7.6_{-2.5}^{+3.3}$ & $22.5_{-4.6}^{+5.3}$ & 0.99/504 \\
35 & $10.18_{-0.84}^{+0.79}$ & $0.17_{-0.01}^{+0.03}$ & $0.81_{-0.06}^{+0.08}$ & $\cdots$ & $0.09_{-0.02}^{+0.03}$ & $0.25_{-0.04}^{+0.08}$ & $0.25_{-0.06}^{+0.09}$ & $\cdots$ & $0.30_{-0.15}^{+0.23}$ & $10.4_{-5.0}^{+5.8}$ & $18.0_{-4.8}^{+3.8}$ & 0.93/283 \\
37 & $5.9_{-4.6}^{+2.8}$ & $0.54_{-0.10}^{+0.21}$ & $\cdots$ & $\cdots$ & $0.21_{-0.14}^{+0.28}$ & $0.28_{-0.15}^{+0.26}$ & $0.41_{-0.21}^{+0.27}$ & $\cdots$ & $<0.27$ & $\cdots$ & $1.88_{-0.99}^{+1.44}$ & 1.06/33 \\
38 & $3.81_{-0.57}^{+0.64}$ & $0.65_{-0.05}^{+0.04}$ & $\cdots$ & $\cdots$ & $0.35_{-0.13}^{+0.16}$ & $0.52_{-0.15}^{+0.19}$ & $0.46_{-0.14}^{+0.16}$ & $\cdots$ & $0.11_{-0.02}^{+0.03}$ & $\cdots$ & $1.46_{-0.25}^{+0.34}$ & 1.12/106 \\
39 & $7.76_{-0.50}^{+0.52}$ & $0.14_{-0.01}^{+0.01}$ & $0.45_{-0.03}^{+0.04}$ & $0.11_{-0.04}^{+0.07}$ & $0.09_{-0.02}^{+0.04}$ & $0.25_{-0.06}^{+0.09}$ & $0.60_{-0.21}^{+0.34}$ & $\cdots$ & $\cdots$ & $10.3_{-2.7}^{+2.7}$ & $66.2_{-25.0}^{+38.1}$ & 1.21/230 \\
40 & $7.82_{-0.53}^{+0.53}$ & $0.13_{-0.01}^{+0.01}$ & $0.42_{-0.03}^{+0.04}$ & $0.07_{-0.03}^{+0.04}$ & $0.08_{-0.03}^{+0.04}$ & $0.12_{-0.04}^{+0.05}$ & $0.82_{-0.25}^{+0.30}$ & $\cdots$ & $\cdots$ & $8.9_{-2.4}^{+2.5}$ & $126.0_{-46.1}^{+69.1}$ & 1.17/244 \\
41 & $6.48_{-0.28}^{+0.29}$ & $0.13_{-0.01}^{+0.01}$ & $0.63_{-0.03}^{+0.03}$ & $0.17_{-0.04}^{+0.04}$ & $0.09_{-0.02}^{+0.02}$ & $0.22_{-0.04}^{+0.06}$ & $0.31_{-0.06}^{+0.08}$ & $\cdots$ & $\cdots$ & $7.9_{-1.7}^{+1.6}$ & $24.9_{-5.3}^{+6.8}$ & 1.11/299 \\
42 & $10.57_{-0.68}^{+0.63}$ & $0.20_{-0.01}^{+0.02}$ & $>1.53$ & $0.13_{-0.06}^{+0.11}$ & $0.33_{-0.08}^{+0.09}$ & $0.37_{-0.10}^{+0.16}$ & $0.32_{-0.08}^{+0.13}$ & $\cdots$ & $\cdots$ & $69.4_{-6.9}^{+6.4}$ & $21.4_{-7.0}^{+8.8}$ & 1.34/170 \\
43 & $7.1_{-1.2}^{+1.4}$ & $0.69_{-0.05}^{+0.05}$ & $\cdots$ & $\cdots$ & $<0.26$ & $0.68_{-0.17}^{+0.20}$ & $0.38_{-0.10}^{+0.11}$ & $\cdots$ & $0.06_{-0.03}^{+0.06}$ & $\cdots$ & $1.06_{-0.17}^{+0.23}$ & 0.89/123 \\
45 & $6.0_{-1.4}^{+7.6}$ & $0.49_{-0.20}^{+0.11}$ & $\cdots$ & $0.39_{-0.39}^{+1.09}$ & $0.12_{-0.09}^{+0.23}$ & $0.29_{-0.13}^{+0.27}$ & $0.16_{-0.15}^{+20.9}$ & $\cdots$ & $0.08_{-0.04}^{+0.09}$ & $\cdots$ & $1.80_{-0.84}^{+5.00}$ & 1.02/54 \\
46 & $13.0_{-2.7}^{+2.9}$ & $1.37_{-0.43}^{+0.93}$ & $\cdots$ & $\cdots$ & $<0.46$ & $0.76_{-0.19}^{+0.26}$ & $0.71_{-0.16}^{+0.22}$ & $\cdots$ & $\cdots$ & $7.9_{-3.9}^{+14.8}$ & $0.18_{-0.08}^{+0.14}$ & 1.02/132 \\
\end{longtable}
}
}

\begin{figure*}[ht]
    \centering
    \includegraphics[width=\textwidth]{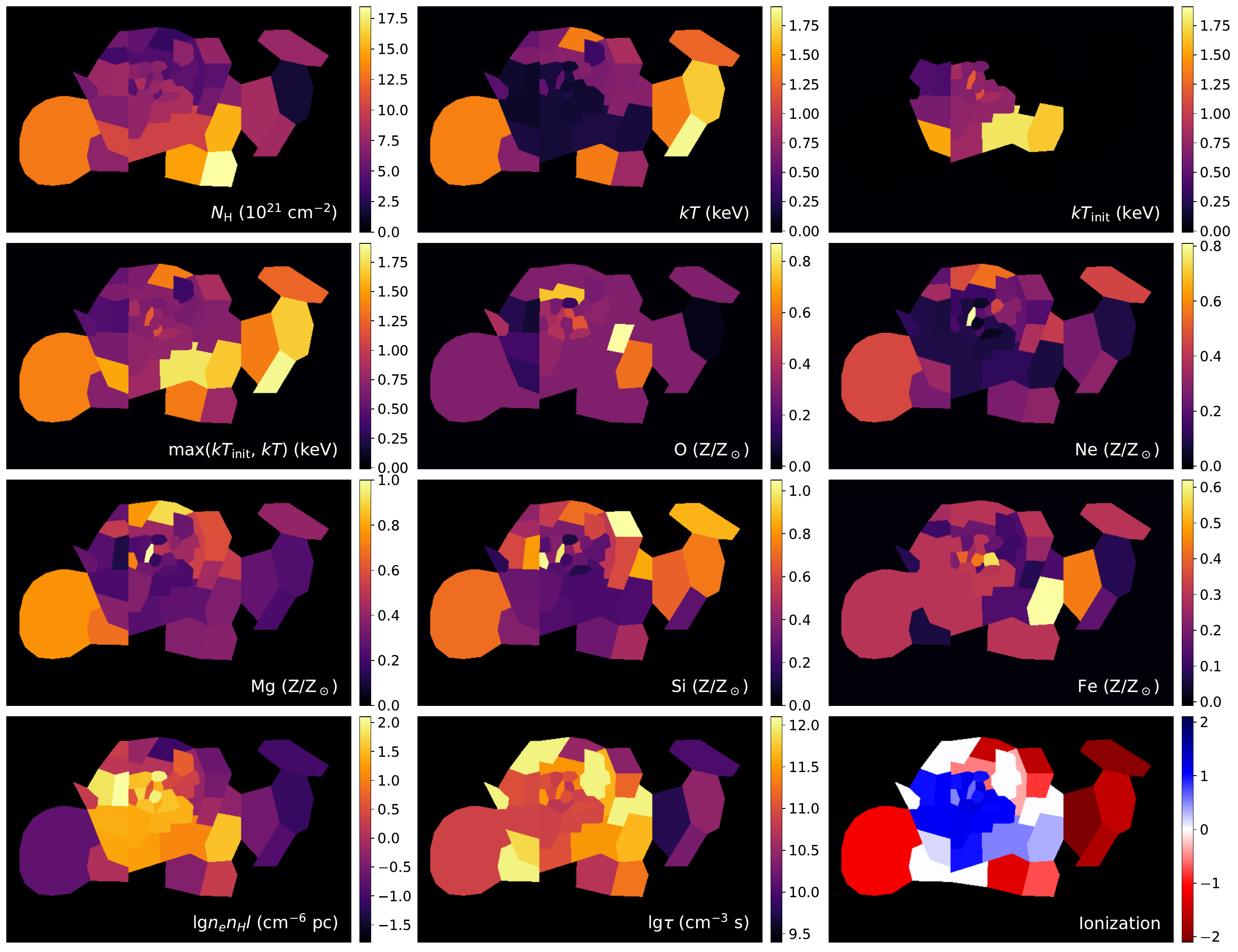}
    \caption{Maps of the fitting parameters with a subsolar abundance Z=0.3\,Z$_\odot$. The value of the parameters and their uncertainty correspond to Table~\ref{tab: fit3}.}
    \label{fig: resultmap3}
\end{figure*}

With a given temperature, the intensity of X-ray emissions from coronal plasma is in proportion to the emission measure (EM). In XSPEC, its definition is
\begin{equation}
    norm=10^{-14}\frac{\int n_\mathrm{e}n_\mathrm{H}\mathrm{d}V}{4\pi D^2},
    \label{eq: emdef}
\end{equation}
where the electron density $n_\mathrm{e}$ is about 1.2 times the proton density $n_\mathrm{H}$ for solar abundance plasma with fully ionized H and He, $D$ is the distance and $V$ is the filling volume, which cannot be directly measured by spectroscopy. Assuming a uniform density, we can adapt the Equation~\ref{eq: emdef} into $EM=n_\mathrm{e}n_\mathrm{H}l=4\pi\times10^{14}norm/\Sigma$. The angular area $\Sigma$ can be transferred from the {\it backscal} keyword. For FXT, $backscal=1$ corresponds to $600\times600$ pixels with an angular size of $9.67''$. Therefore, we can give a direct measurement independent of the real geometry and the distance

\begin{equation}
    EM=n_\mathrm{e}n_\mathrm{H}l=(0.515\,\mathrm{cm^{-6}\,pc})\,\frac{norm}{\Sigma}.
\end{equation}

The EM is more sensitive to the density than the ``depth'' $l$ associated with the geometry. The surface brightness of W28 generally follows a centrally-filled morphology, with a smooth variation of $l$ in the SNR interior \citep[see Figure~1 in][]{Rho98}. The other structures in the east or west, if associated with W28, are expected to have lengths of a similar order of magnitude in the line of sight. 

The fitting parameters are listed in Table~\ref{tab: fit3} for all the regions except Region 44 (see Table~\ref{tab: cal}), the northeast shell mentioned above, and Region 36 due to weak X-ray emissions there. For several regions that can be fitted by more than one thermal component sets with different ionization states or metal abundances, all the results are maintained for reference. Fig.~\ref{fig: resultmap3} maps the spatial distribution of different parameters except for the S abundances, which can only be constrained in a few regions. The parameters fixed at default values were set as blank, and only the ``conservative model'' with less physical hypothesis is reserved here. For example, if two sets of model parameters give similar fitting statistics ($\chi_r^2$), the one with solar or subsolar abundance is preferred over the supersolar one, since most regions in this mature SNR show subsolar abundance.

Fig.~\ref{fig: resultmap3} nicely reflects how different parameters vary among different regions. The absorption changes slightly around $\sim8\times10^{21}$\,cm$^{-2}$ in most areas of W28 center, but rises from the north to the south, peaking up to $\sim1.8\times10^{22}$\,cm$^{-2}$ at Region 10, a clumpy structure southwest of the center. No absorption enhancement is found associated with the molecular clouds (Regions 20, 38, and 45), suggesting that these clouds may be located behind the SNR.

For metals, our result generally supports a subsolar abundance of O, Ne, Mg, Si, and Fe. With current data, spectra of most regions cannot give a good constraint of O and Fe. Also, a strong degeneracy between the initial temperature and Fe abundance appears in some regions, especially the southwest part of the W28 center. We found that O and Fe seem to peak at the SNR center and decay near the edge. On the contrary, the abundance of Ne, Mg, and Si is enhanced on the edge, especially in the north. However, it should be noted that the uncertainties of these metals are large ($\gtrsim25\%$) with current data. 

We also added two panels for a better display in Fig.~\ref{fig: resultmap3}. To underline the ionization state varying among different regions, we define a parameter named ``ionization degree''to reflect the deviation from the CIE. Its absolute value equals $-\lg (\tau/(10^{12}\,\mathrm{cm^{-3}\,s}))$ on the assumption that plasma with $\tau\geq10^{12}\,\mathrm{cm^{-3}\,s}$ can be considered to reach CIE. Its sign would be either positive (coded in red) for recombination or negative (coded in blue) for ionization. We note that the NEI plasma slightly deviate from CIE even when $\tau>10^{12}\,\mathrm{cm^{-3}\,s}$. This deviation is stronger in recombining plasma, which radiates strong radiative-recombining continuum (RRC) and the spectral shape could still significantly differ from the CIE case.

\begin{figure}[ht]
    \centering
    \subfigure{\includegraphics[width=\linewidth]{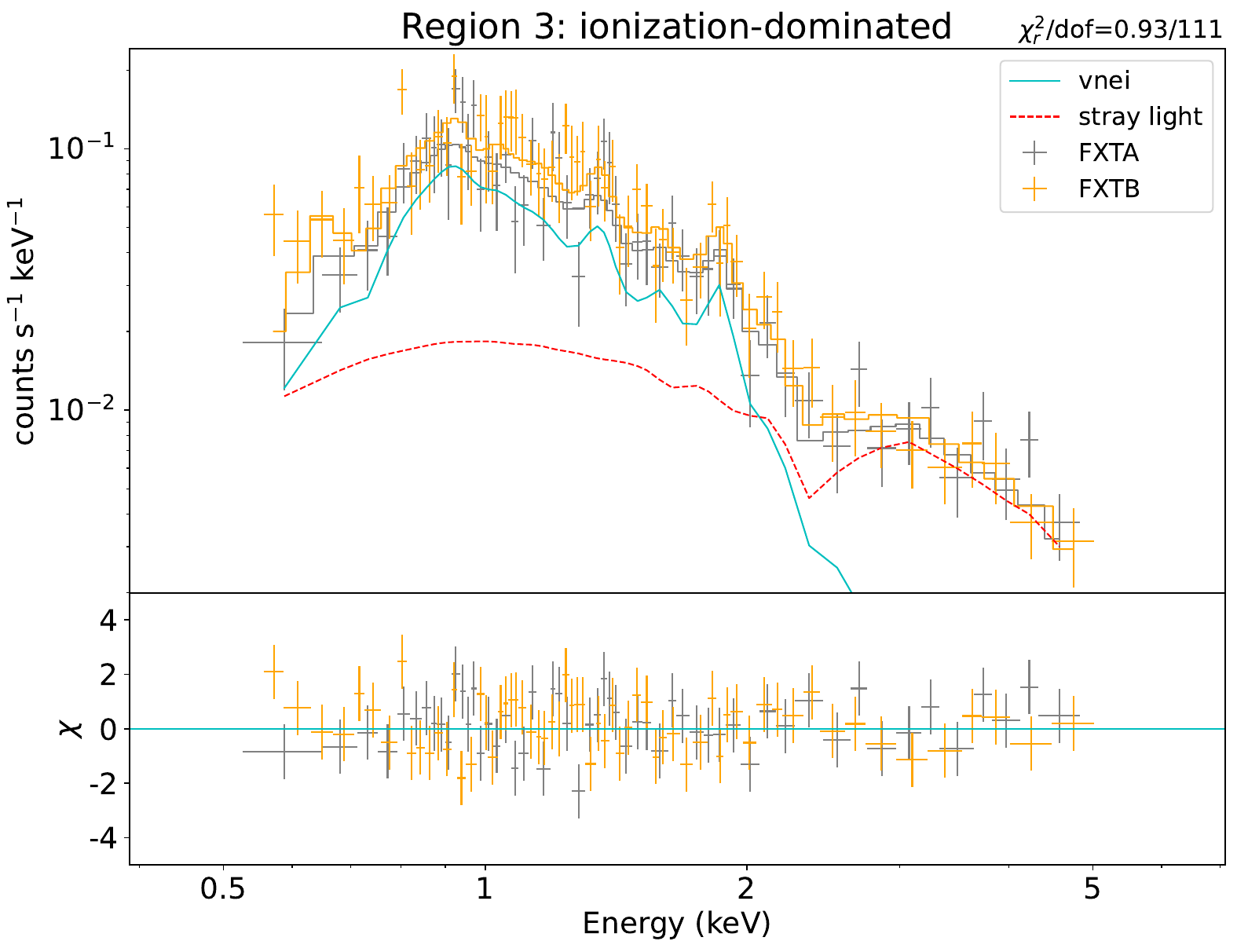}}
    \subfigure{\includegraphics[width=\linewidth]{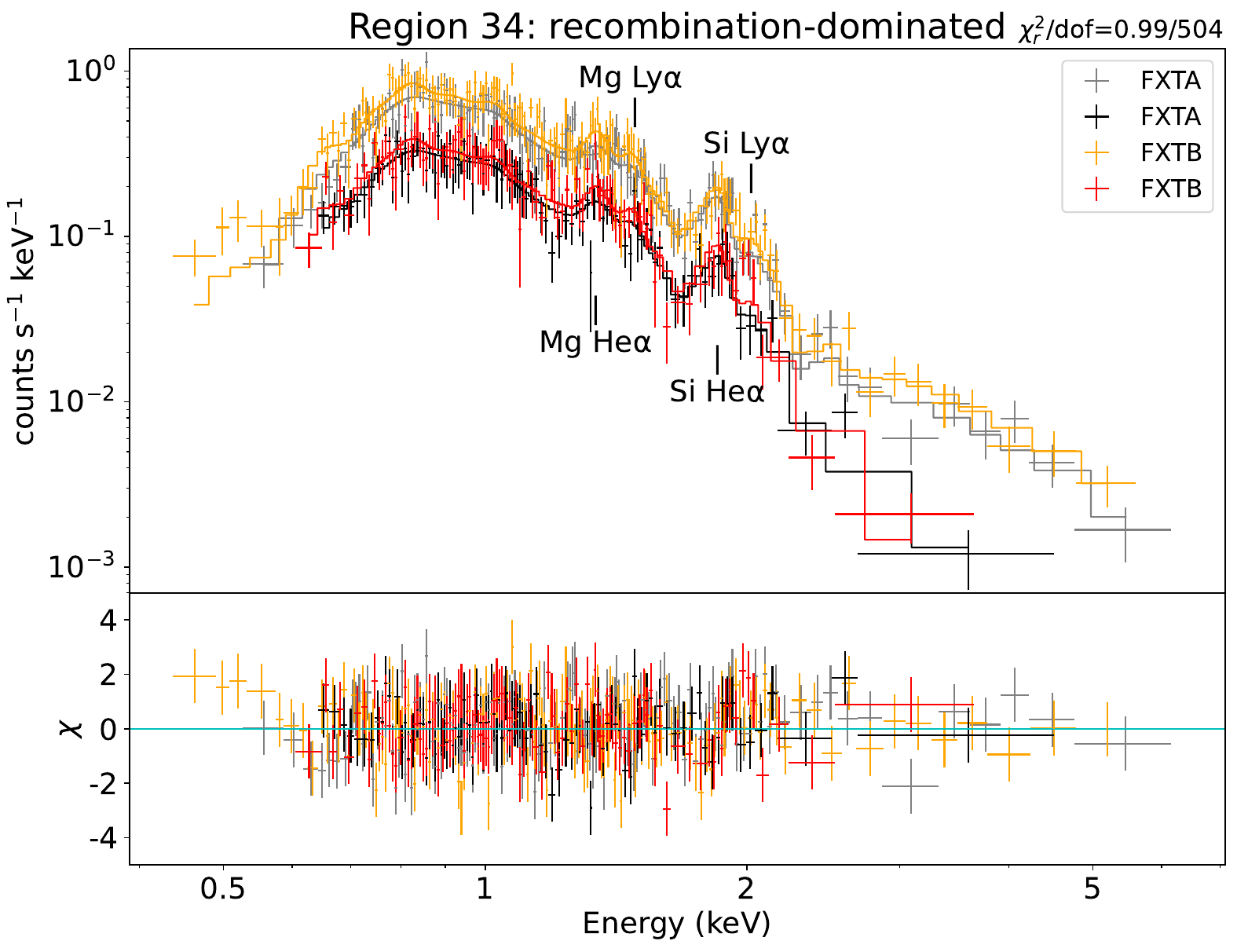}}
    \caption{Examples of spectra of underionized plasma (upper panel, Region 3 in the western shell) and overionized plasma (lower panel, Region 34 in the W28 center). In the upper panel, the contribution of the plasma and the stray light are marked with cyan solid lines and red dashed lines, respectively. The lower panel contains spectra of two observations, of which only the yellow and grey data points are significantly affected by the stray light.}
    \label{fig: specexample}
\end{figure}

In previous studies, recombination plasma has been discovered in the W28 center and the southwest clump (Region 9) \citep{Sawada12, Himono23}. However, our maps give a more complex picture. All the recombination-dominated regions are located at the center and southeast of the center, with electron temperatures around 0.2\,keV and initial temperatures around 0.7\,keV. However, there is a temperature gradient from the southwest (peaks at $\sim1.8$\,keV) to the northeast, as previously reported by \citet{Rho02}. The EM in recombining regions are significantly higher than the other regions, indicating a higher density. On the contrary, in the north of W28, the plasma tends to be underionized. The boundary between the ionizing and recombining plasma is coincident at the molecular clouds (Regions 20 and 38) and the inner edge of the northern radio shell (Regions 16, 17, 20, and 38). 

The difference in the ionization state can also be well reflected by the spectral shape. Fig.~\ref{fig: specexample} shows typical under- and over-ionization-dominated spectra in our analysis (Region 3 and 34, respectively). Although Region 3 has a much higher temperature of $\sim1.9$\,keV, it emits lines with lower ionization levels, like \ion{Ne}{IX} near 0.9\,keV, \ion{Mg}{XI} near 1.3\,keV, and \ion{Si}{XIII} near 1.8\,keV. On the contrary, lines with higher ionization states also appear in Region 34 dominated by recombination, showing \ion{Ne}{X} near 1.0\,keV, \ion{Mg}{XII} near 1.5\,keV, and \ion{Si}{XIV} near 2.0\,keV. These H-like lines reflect a high ionization temperature as reported in \citet{Kawasaki05, Sawada12}. The \ion{Mg}{XII} RRC can contribute near $\sim2$\,keV as well. 

We also define a temperature $\max(kT_\mathrm{init}, kT)$, which equals the electron temperature CIE or ionization-dominated regions but the initial temperature in recombining regions. This parameter is to test the hypothesis that the gas temperature in W28 was initially near uniform before undergoing localized and recent recombination processes. Under this hypothesis, the initial temperature of the recombining region is similar to the temperature of nearby regions close to CIE \footnote{The {\it vrnei} model describes the recombination of the plasma in CIE initially. However, the real picture could be that the plasma is first shocked and heated, and then, before it reaches CIE, the recombination begins. This scenario is so complex that it exceeds the range of our study, and most regions have an ionization timescale of $\sim10^{12}\,\mathrm{cm^{-3}\,s}$, close to CIE.}. As shown in Fig.~\ref{fig: resultmap3}, most regions in the W28 center share a similar $\max(kT_\mathrm{init}, kT)$ of $\sim0.6$--$0.7$\,keV, including the recombining regions with a temperature of $\lesssim0.2$\,keV. On a larger scale, there is a gradient for this temperature, where regions in the outer part of the remnant (e.g. the western shell) shows a higher temperature of $>1$\,keV. A further discussion on the temperature along with the gas density is provided in Sect.\,\ref{sec: phase}.

The boundary regions of W28 show a large variation of spectral properties. We listed them in detail below.

Region 1--4: The western shell. These regions show a high plasma temperature of 1.2--1.9\,keV, in an underionized state with an ionization timescale of $\sim10^{10}$\,s\,cm$^{-3}$, and a low density (EM). 

Region 9 and 12: These two regions are located in the southwest of W28 but show different properties from the W28 center. Both regions share a high foreground absorption, high initial temperature of $\sim1.7$\,keV compared to $\sim0.7$\,keV in the W28 center, and a long recombining timescale close to CIE, although the current electron temperature are near $\sim0.2$\,keV. The long recombining timescale is also reported by \citet{Himono23}.

Region 10 and 13: The southernmost regions in our spectral analysis, spatially associated with a segment of the radio shell (Fig.~\ref{fig: 3bandrgb}). Region 10 corresponds to Clump E, while Region 13 has a low surface brightness. Both regions are ionization dominated with a higher temperature of $\gtrsim0.8$\,keV, while the plasma north to them are recombining with a initial temperature of $\sim0.7$\,keV. 

Region 46: The diffuse X-ray emissions east of W28 center, sharing a similar size with the radio SNR G6.5$-$0.4 but evidently located by the south. The absorption here is stronger than in neighboring regions. This region has a higher temperature of $\sim1.4$\,keV and the plasma is underionized.

\bigskip
In the end of this section, we compare our result, based on a single-temperature NEI model, with previous studies. Generally, one NEI model can reproduce the plasma in W28 \citep{Kawasaki05, Sawada12, Zhou14, Pannuti17, Okon18, Himono23}, while a dual-temperature model was proposed as well \citep{Rho02, Zhou14, Pannuti17, Okon18, Himono23}. Despite the difference in the specific models, the fitted temperature can be divided in three populations: (1) the coolest one near 0.2--0.4\,keV \citep{Sawada12, Zhou14, Okon18, Himono23}, corresponding to our electron temperature here; (2) a warm temperature near 0.6--0.7\,keV \citep{Rho02, Zhou14, Pannuti17}, similar to our initial temperature; and (3) a hot one above 1\,keV \citep{Rho02, Kawasaki05, Zhou14, Pannuti17}, either as the initial temperature \citep{Zhou14, Sawada12, Pannuti17, Okon18, Himono23}, or as a hot component \citep{Rho02, Kawasaki05, Zhou14, Pannuti17}. Here the hot temperature above 1\,keV is ``missing'' in our study, just as the comparison between XMM-Newton and FXT in the NE shell. Due to short exposure, finer region sizes, and effect of stray light, the spectrum quality above $\sim2$\,keV is not good enough for a search and analysis for a harder/hotter component. Our spectral analysis only focuses on the data below $\sim2$\,keV, which dominates the X-ray flux of W28.

\section{Discussion}
\label{sec: dis}
\subsection{Temperature-EM diagram}
\label{sec: phase}
The X-ray observations reveal a variation of plasma properties across SNR W28. A comparison of the density and temperature relation among regions can provide insight into the physical origin of this variation. For example, a random distribution may suggest no physical relations, while a power-law relation $EM\sim T^\alpha$ can reflect the dynamical evolution of the plasma. A positive index $\alpha\approx 3$ (i.e., $n\sim T^{3/2}$) indicates adiabatic expansion, while a negative index $\alpha\approx-2$ (i.e., $n\sim T^{-1}$) may infer an isobaric scenario \citep{Zhang19}. The power-law relation holds not directly for the density and temperature themselves but for their changes compared to the initial value. Here, uniform temperature and density are assumed in the early stage of the SNR.

\begin{figure}
    \centering
    \includegraphics[width=0.495\textwidth]{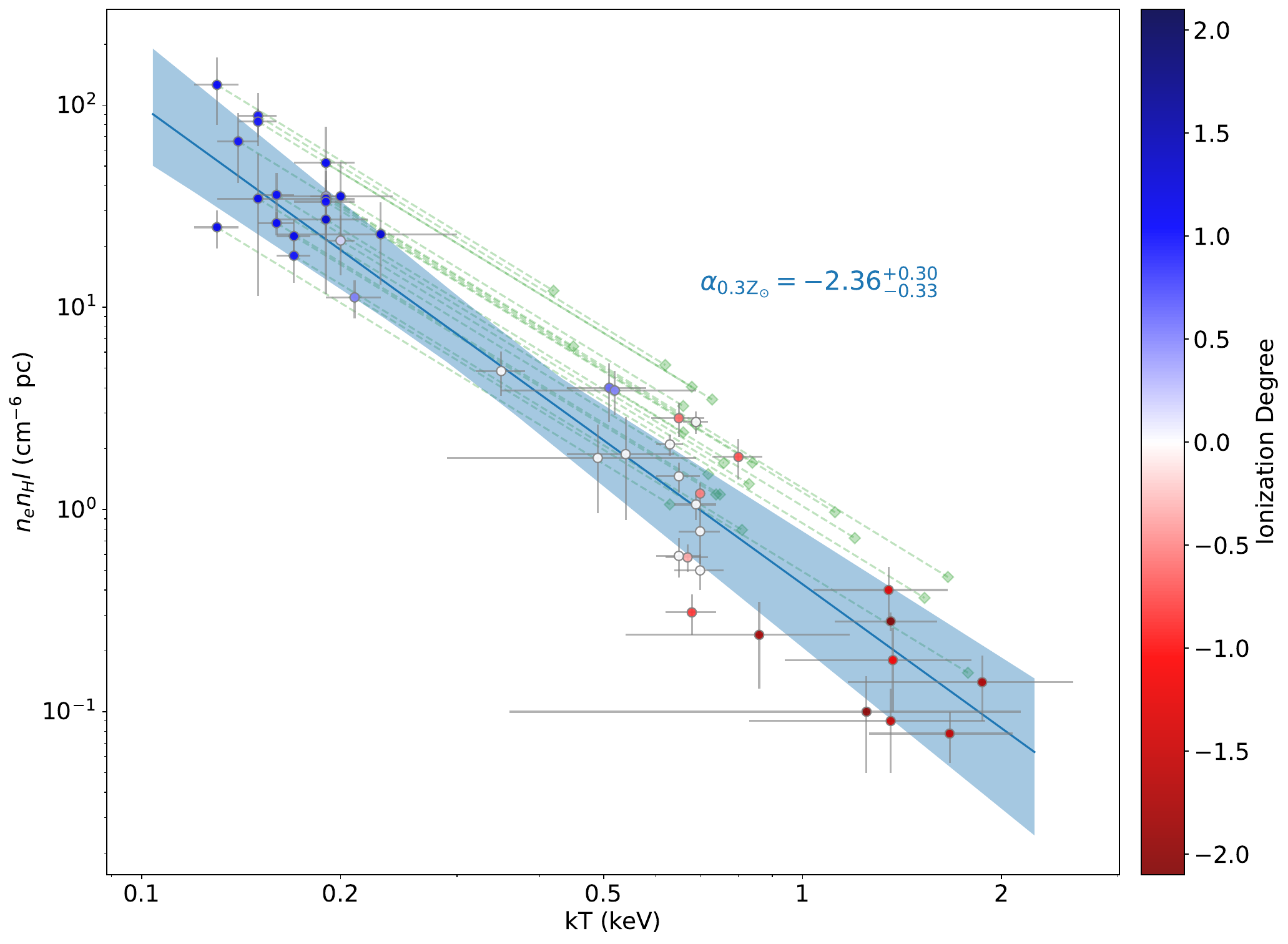}
    \caption{The temperature-EM of the best-fit value of the EM and the temperature (circle dots). The blue solid line is the best-fit powerlaw model and the shadow is the 90\% confidence belt. Green diamonds and dashed lines mark the ``initial'' property of the recombining plasma on the assumption of an isobaric cooling. The color of the markers shows the ionization degrees, which is the same as Fig.~\ref{fig: resultmap3}.}
    \label{fig: phase3}
\end{figure}

Figure ~\ref{fig: phase3} plots the relation between the best-fit values of EM ($=n_\mathrm{e} n_{\rm H} l$) and $kT$ (Fig.~\ref{fig: phase3}). We use EM instead of the density since it is a direct measurement that is proportional to the fitted parameter $norm$, but the density of the plasma is degenerated with the assumed geometry.

The EM-$kT$ distribution can be divided into three different populations. In terms of the temperature, the coolest population is around $\sim0.1$--$0.2$\,keV and the gas is recombining. Most of these regions are distributed in the SNR interior. On the contrary, the population with a temperature over 1\,keV are underionized and are mostly located in the western shell and the southeastern lobe (region 46). The rest mainly have a temperature around 0.6--0.7\,keV with various ionization degrees. However, the global distribution of these data points can be described by a power-law function, and the scatters generally fall within 1 dex.

The fitted power-law index $\alpha$ of $-2.36_{-0.33}^{+0.30}$ (90\% statistical error) is probably indicative of the isobaric scenario where $\alpha=-2$ (i.e., $nT=$ const.). Using the fitting result with the solar abundance, this isobaric correlation still exists with an index of $\sim-2.27$ with relatively large scatter (see Appendix~\ref{app: fitsolar}). On a large scale, the plasma in W28 follows the near-isobaric relation. The western shell, originally identified as distinct SNRs, conforms to this relationship and may therefore belong to W28.

For recombination dominated by thermal conduction, the gas is expected to experience an isobaric process. We mark the ``initial'' state of the recombining regions in Fig.~\ref{fig: phase3} (green diamonds) assuming an isobaric heat exchange (i.e., $\alpha=-2$). The distribution of the initial temperatures, derived from the fitted $kT_\mathrm{init}$ in Table~\ref{tab: fit3}, is mostly consistent with the 0.6--0.7\,keV population, although the initial EMs are seemingly higher within $\sim0.5$\,dex. The 0.7-keV population dominant in the remnant center can be referred as the initial state of W28 gas before the cooling.

The absence of a strong pressure gradient rules out an efficient ongoing adiabatic cooling, even though the present isobaric conditions might have resulted from the relaxation of an earlier gradient. Simulations of thermal-composite SNRs reveal that the recombination due to adiabatic cooling is expected to take place near the shock interface rather than associated with cloudlets \citep[see region ``b'' and ``d'', respectively, in Figure 6 \& 7 by ][]{Zhang19}. In W28, the recombining regions show a strong spatial coincidence with H$\alpha$ emission. This correspondence favors a thermal conduction origin over an adiabatic-cooling origin for the recombining plasma.

Compared to the temperature, the EM distribution shows a larger scatter, which can more or less account for the dependence on the geometry of the density. For SNRs, the radial distribution of the gas $l$ can be roughly estimated under different scenarios. One is that the X-ray-emitting gas is swept into a thin shell adiabatically with a depth of $\sim1/12$ of the remnant radius. The other extreme case corresponds to the hot gas uniformly filled in the entire SNR. Therefore, the uncertainty of $l$ is expected within $\sim1$\,dex, which leads to the uncertainty of the density only within $\sim3$ times according to the definition of the EM. Moreover, the real spatial distribution of $l$ should be continuous in structures with smooth surface brightness, such as the W28 center. 

W28 has a radius of $\sim20'$ in the radio band, corresponding to $\sim11$\,pc at a distance of 1.9\,kpc \citep{Velazquez02}. Therefore, the radial depth $l$ can be $\sim1$--$22$ parsecs. Considering the uncertainty of the geometry, the density is $\sim0.1$--$0.4$, $\sim0.3$--$1.5$, and $\sim2$--$7$\,cm$^{-3}$ for the western shell, W28 center without recombination, and the recombining gas, respectively. This density gradient of W28 decreasing from the east to the west may account for the blow out morphology of the western shell.

\subsection{Origin of the recombining plasma}
There has been no consensus on the origin of recombining plasma in SNRs, as it depends on the specific conditions of the SNR and the environment in which it evolves.
Thermal conduction \citep{Cox99, Shelton99, Zhou14}, cloud evaporation \citep{Cowie81, White91, Long91}, adiabatic cooling \citep{Itoh89, Zhang19}, or other mechanisms like low-energy cosmic rays \citep{Himono23} are potential origins. With clumpy H$\alpha$ emissions in the center, which show different morphology from the radiative shocks \citep{Long91, White91}, it is natural to consider the thermal conduction and cloud evaporation to play an important role in the cooling.

From the spectral analysis, the recombination (or cooling) timescale can be roughly estimated: 
\begin{equation}
    t_\mathrm{rec}=\tau/n_\mathrm{e}=3.2\,\mathrm{kyr}\,\left(\frac{\tau}{10^{11}\,\mathrm{s\,cm^{-3}}}\right)\,\left(\frac{n_\mathrm{e}}{1\,\mathrm{cm^{-3}}}\right)^{-1},
\end{equation}
which is consistent with most areas in W28 center according to a previous study according to \citet{Himono23}. This timescale indicates that the cooling took place recently compared to the dynamical timescale of $\sim33$--42\,kyr proposed before \citep{Rho02, Velazquez02, Li10} and $\sim8
$\,kyr suggested in our study (see the subsection below). It should be noted that $t_\mathrm{rec}$ may not be valid for the adiabatic cooling, where the density can drops rapidly.

Previous studies mainly focused on the long-range/classical thermal conduction \citep{Zhou14, Himono23}. The density gradient (Fig.~\ref{fig: resultmap3}) from east to west may support such a scenario that the molecular clouds in the east can be a cooling source. The long-range means the conduction scale exceeds the mean free path of the electron \citep{Cowie77}
\begin{equation}
    \lambda_\mathrm{e}=0.1\,\mathrm{pc}\,\left(\frac{n_\mathrm{e}}{1\,\mathrm{cm^{-3}}}\right)^{-1}\,\left(\frac{kT_\mathrm{e}}{0.7\,\mathrm{keV}}\right).
\end{equation}
The conduction timescale is
\begin{equation}
    t_\mathrm{con}=\frac{kn_\mathrm{e}l_\mathrm{T}^2}{\kappa}=(38\,\mathrm{kyr})\,\left(\frac{n_\mathrm{e}}{1\,\mathrm{cm^{-3}}}\right)\,\left(\frac{l_\mathrm{T}}{10\,\mathrm{pc}}\right)^2\,\left(\frac{kT}{0.7\,\mathrm{keV}}\right)^{-\frac{5}{2}}\,\left(\frac{\ln\Lambda}{32}\right),
\end{equation}
where $\kappa$ is the thermal conductivity as a function of temperature \citep[see Equation 3--5 in][]{Cowie77} and $\ln\Lambda=32+\ln (n_\mathrm{e}/1\,\mathrm{cm^{-3}})^{-1}(kT/0.7\,\mathrm{keV})$ is the Coulomb logarithm. $l_\mathrm{T}=T/\nabla T$ is the scale length of the temperature gradient, which can be considered as the size of the recombination region \citep{Zhou14, Himono23}. The long-range cooling timescale $t_\mathrm{con}$ is far longer than the recombination timescale of $\sim3$\,kyr, and the magnetic field can suppress the conduction. In other words, the long-range cooling may contribute but may only affect a small part of the SNR. 

Therefore, we consider if the small-scale saturated thermal conduction works efficient in W28. One crucial indication is the clumpy H$\alpha$ emissions in the remnant center \citep{White91, Long91}, which is found highly spatially correlated with the recombining regions in our analysis mentioned in the section above. One explanation of the H$\alpha$ emission is that the gas evaporated from the cloud is heated to $\sim10^4\,$K \citep{White91, Zhang19}. In numerical simulations, the temperature of most cloudlets can increase rapidly to $\sim10^4\,$K behind the shock \citep{Zhang19} after engulfed by the shock. The clouds emitting H$\alpha$ emission are likely in the interior of the SNR rather than on the SNR shell, since they do not spatially coincide with any radio, infrared or molecular structures. 

We herein propose a toy model to briefly calculate the effect of saturated thermal conduction \citep{Cowie77}, where the heat flux is
\begin{equation}
    F_\mathrm{sat}=0.4\left(\frac{2kT_\mathrm{e}}{\pi m_\mathrm{e}}\right)^\frac{1}{2}n_\mathrm{e}kT_\mathrm{e}.
\end{equation}
Assuming the heat exchange happens at the interface between the cloud and the hot gas, the heat transfer rate per unit volume to cool the hot gas is

\begin{equation}
    \dot{u}=-4\pi r_\mathrm{c}^2N_\mathrm{c}F_\mathrm{sat}=\frac{3}{2}n_\mathrm{e}k\dot{T_\mathrm{e}},
    \label{eq: heat}
\end{equation}

where $r_\mathrm{c}$ is the average radius of the cloudlet and $N_\mathrm{c}$ is the number density of cloudlets. If we do not consider the variation of the cloud sizes, integrate the equation and we can get the evolution of $T_\mathrm{e}$ under the saturated thermal conduction

\begin{equation}
    \begin{split}
        t_\mathrm{sat} &=\frac{15}{8\pi r_\mathrm{c}^2N_\mathrm{c}}\left[\left(\frac{2kT_\mathrm{e}}{\pi m_\mathrm{e}}\right)^{-\frac{1}{2}}-\left(\frac{2kT_\mathrm{init}}{\pi m_\mathrm{e}}\right)^{-\frac{1}{2}}\right] \\
        & =(5.7\,\mathrm{kyr})\,\left(\frac{r_\mathrm{c}}{0.1\,\mathrm{pc}}\right)^{-2}\left(\frac{N_\mathrm{c}}{1\,\mathrm{pc}^{-3}}\right)^{-1},
    \end{split}
\end{equation}
if $kT_\mathrm{e}=0.2$\,keV and $kT_\mathrm{init}=0.7$\,keV. Considering the mass loss of the cloudlets, the saturated thermal conduction timescale would be longer but still comparable to the recombination timescale with a higher specific surface area (i.e., higher $r_\mathrm{c}^2N_\mathrm{c}$). Therefore, the warm gas with a more diffuse distribution may play a more crucial role in the cooling than the denser cooler molecular clouds with a similar mass. In the H$\alpha$ image, the scale size of the clumps is $\sim10''$, or $\sim0.1$\,pc at 1.9\,kpc, which generally equals the mean free path of the electrons, indicative of saturated thermal conduction.

The cloud evaporation takes place as the cold gas is heated by the saturated thermal conduction. The density, or rather, the EM in the recombining areas is several times higher than those in CIE or underionization (see Fig.~\ref{fig: phase3}), suggesting that the X-ray-emitting gas therein is significantly enriched through the cloudlet evaporation. Under a rough estimation based on the EM measured in the spectral analysis, the evaporated gas may contribute $\sim50$--$60$\,M$_\odot$ mass. 

Our X-ray spectral analysis (Sect.\,\ref{sec: spec}) reveals consistently subsolar metal abundances throughout W28, present in both the recombining regions dominated by evaporated cloud material and the others with the shocked low-density intercloud medium (ICM). This would be unusual given its location in the inner Galactic Plane (G6.4$-$0.1), where near-solar metallicity is generally expected since metallicity typically decreases with increasing Galactocentric distance. However, it cannot be easily precluded as some H\,II regions and planet nebulae exhibit a [O/H] abundance 0.5--1\,dex below this gradient \citep{Maciel09, Wenger19}. The subsolar abundance of neon, an element largely inert to dust depletion, argues against dust depletion being the primary cause. We also cannot rule out the possibility that an extra continuum component affects the abundance measurements. Thus, the apparently low metal abundance in W28, a feature also noted in other SNRs such as the Cygnus Loop \citep[e.g.,][]{Katsuda08}, remains an open question, likely requiring future multi-wavelength analysis and X-ray missions with high spectral resolution.

We briefly compare the toy model here with the self-similarity model to describe the cloud evaporation \citep{White91} to end this subsection. In their study, the dimensionless parameter $C$ is the mass ratio between the cloudlets and ICM, and $\bar{\tau}$ (using a bar marker to distinguish it from the ionization timescale) is the ratio of the timescale the cloud depletes through evaporation (evaporation timescale $t_\mathrm{evap}$) to the age, or the dynamical timescale $t_\mathrm{dyn}$ here, of the SNR. According to its definition, we can derive
\begin{equation}
    C=\frac{m_\mathrm{c}}{m_0}=\frac{4\pi n_\mathrm{c}N_\mathrm{c}r_\mathrm{c}^3}{3n_0}=4.2\,\left(\frac{n_\mathrm{c}/n_0}{10^3}\right)\,\left(\frac{N_\mathrm{c}}{1\,\mathrm{pc}^{-3}}\right)\,\left(\frac{r_\mathrm{c}}{0.1\,\mathrm{pc}}\right)^3,
\end{equation}
where $n_\mathrm{c}$ is the density in the cloudlet and $n_0$ is the pre-shock ICM density. \citet{Long91} derived a density of $\sim25$\,cm$^{-3}$ from the [\ion{S}{II}] doublet ratio, which corresponds to a temperature of $\sim10^4$\,K, and the cloudlet would be even denser with a lower temperature. Along with the shocked gas density of $\sim1$\,cm$^{-3}$ without recombination in the W28 center, we estimate the lower limit of $n_\mathrm{c}/n_0$ is around 100. On the other hand, the Equation~\ref{eq: heat} can be rewritten to describe the mass loss as
\begin{equation}
    \dot{U}=4\pi r_\mathrm{c}^2F_\mathrm{sat}=-4\pi r_\mathrm{c}^2n_\mathrm{c}\dot{r_\mathrm{c}}\,\frac{3}{2}kT_\mathrm{e}.
\end{equation}
Thus the evaporation timescale can be described as the ratio of cloudlet size $r_\mathrm{c}$ to its rate:
\begin{equation}
\begin{split}
    t_\mathrm{evap} & =\left|\frac{r_\mathrm{c}}{\dot{r_\mathrm{c}}}\right|=\frac{15}{4}r_\mathrm{c}\,\left(\frac{\pi m_\mathrm{e}}{2kT_\mathrm{e}}\right)^\frac{1}{2}\frac{n_\mathrm{c}}{n_\mathrm{e}} \\
    & =(16\,\mathrm{kyr})\,\left(\frac{kT_\mathrm{e}}{0.2\,\mathrm{keV}}\right)^{-\frac{1}{2}}\,\left(\frac{n_\mathrm{c}/n_0}{10^3}\right)\,\left(\frac{r_\mathrm{c}}{0.1\,\mathrm{pc}}\right),
\end{split}
\end{equation}
where $n_\mathrm{e}=4.8n_0$ assuming the electron density corresponds to the shocked ICM with fully ionized H and He. A larger $\bar{\tau}=t_\mathrm{evap}/t_\mathrm{dyn}\gtrsim1$ is favored for the cloudlet evaporation \citep{White91}, which may suggest a shorter dynamical timescale of $\sim8$\,kyr (see the subsection below) instead of $\sim33$--42\,kyr proposed in previous studies \citep{Rho02, Velazquez02, Li10}, although the cloudlet density is of high uncertainty. We can get the relation between $C$ and $\bar{\tau}$ under the saturated conduction, which is independent of the density ratio:
\begin{equation}
    \frac{C}{\bar{\tau}}=2.1\,\left(\frac{kT_\mathrm{e}}{0.2\,\mathrm{keV}}\right)^\frac{1}{2}\,\left(\frac{t_\mathrm{dyn}}{8\,\mathrm{kyr}}\right)\,\left(\frac{N_\mathrm{c}}{1\,\mathrm{pc}^{-3}}\right)\,\left(\frac{r_\mathrm{c}}{0.1\,\mathrm{pc}}\right)^2.
\end{equation}
In the model of \citet{White91}, $1\lesssim\bar{\tau}\lesssim C$ is needed to allow the cloudlet evaporation to be efficient and show a morphology deviating from that predicted by the Sedov-Taylor solution. 

\subsection{The western shell and its physical connection to W28}
The most evident newly discovered structure revealed by the large field of view of our EP-FXT observations is the shell-like structure in the west of W28. Currently, these observations support its shock origin including

\begin{itemize}
    \item Shell-like morphology in radio, optical, and X-ray bands (Sect.\,\ref{sec: mor});
    \item Under-ionized hot plasma (Sect.\,\ref{sec: spec}), enhanced [\ion{S}{II}]/H$\alpha$, and non-thermal radio index (see references in Sect.\,\ref{sec: mor}).
\end{itemize}

Although further observations are needed to definitely establish the relationship between this structure and W28, some evidence supports the hypothesis that the western shell is part of the SNR W28:

\begin{itemize}
    \item All of these shells, including the newly discovered western shell and previously known southwestern shell (Clumps D and E), share similar rather than random orientations, which is reasonable for these shells to originate from an explosion near the W28 center (Fig.~\ref{fig: 3bandrgb});
    \item These shells share similar foreground absorption with W28 center (Sect.\,\ref{sec: spec});
    \item The shells and the W28 center are generally isobaric (Sect.\,\ref{sec: phase});
    \item If the western shell belongs to W28, its dynamical age matches the Clump E to the southwest (see details below);
\end{itemize}

Using the parameters of the shell-like structures, the dynamical age and explosion energy can be roughly estimated by the Sedov-Taylor solution.
For non-relativistic strong shock, its velocity depends on its temperature
\begin{equation}
    v_\mathrm{s}=\sqrt{\frac{16kT}{3\mu m_\mathrm{p}}}=(923\,\mathrm{km\,s}^{-1})\,\left(\frac{kT}{1\,\mathrm{keV}}\right)^\frac{1}{2},
    \label{eq: velo}
\end{equation}
where the electron and ion are assumed to share the same temperature and the mean particle mass $\mu=0.6$ for solar abundance, and $m_\mathrm{p}$ is the proton mass. Evolving isotropically and adiabatically in a uniform medium, the dynamical age can be given by 
\begin{equation}
    t_\mathrm{dyn}=\frac{2R_\mathrm{s}}{5v_\mathrm{s}}=(7.0\,\mathrm{kyr})\,\left(\frac{\theta}{30'}\right)\left(\frac{kT}{1\,\mathrm{keV}}\right)^{-\frac{1}{2}}D_{1.9},
\end{equation}
where $D_{1.9}$ is the distance of W28 in units of 1.9\,kpc \citep{Velazquez02} and $\theta$ is the radius of the shell from the center.

For Region 1--3, the temperature is $\sim1.5$\,keV, and the angular distance from the remnant center is $\sim40'$, corresponding to $t_\mathrm{dyn}=7.6$\,kyr. Coincidentally, $t_\mathrm{dyn}=7.8$\,kyr for Region 10 (Clump E) on the southwestern boundary of the radio shell, which is $\sim30'$ away from center and has a fitting temperature of $\sim0.8$\,keV. This dynamical timescale is generally consistent with the ionization timescale of order $\sim10^{10}$\,s\,cm$^{-3}$ in the western shell (assuming $n_\mathrm{e}\sim0.2$\,cm$^{-3}$) and $\sim10^{11}$\,s\,cm$^{-3}$ in the southwest (assuming $n_\mathrm{e}\sim1$\,cm$^{-3}$). This dynamical timescale is reasonably longer than the recombination timescale $t_\mathrm{rec}\sim3$\,kyr. However, it is much shorter than several tens of kilo-years in the previous studies \citep{Rho02, Velazquez02, Li10}, which were based on the velocity of optical emissions (e.g., H$\alpha$) and an assumption of radiative phase. The expansion velocity of $\sim80$\,km\,s$^{-1}$ derived from the optical emission reflects the low shock velocity in the dense medium. 

For SNR evolving in a Sedov-Taylor phase, the explosion energy can also be estimated through
\begin{equation}
    E_\mathrm{exp}=\frac{R^5\rho_0}{\xi t_\mathrm{dyn}^2}=2\times10^{51}\,\mathrm{erg}\,\left(\frac{\theta}{30'}\right)^{3}\,\left(\frac{kT}{1\,\mathrm{keV}}\right)\,\left(\frac{n}{1\,\mathrm{cm^{-3}}}\right)D_{1.9}^3,
    \label{eq: exp}
\end{equation}
where the dimensionless factor $\xi=2.025$, the ambient density $\rho_0=1.4m_\mathrm{p}n/4$. Thus we can roughly estimate the explosion energy of W28 of $\sim(1$--$2)\times10^{51}$\,erg according to the western shell and the southwestern shell. If evolving in a wind bubble, the dynamical age and explosion energy would be $\sim1.5$ times of the values above \citep[see][and references therein]{Chi24}. It should be noted that these calculations are on the assumption that these structures as part of the SNR. If these shells are separate SNRs at different distances, their radius and evolutionary stage could largely deviate from the above results.

W28 is one of the prototype thermal-composite SNRs \citep{Long91} with its morphology deviated from the prediction of the Sedov-Taylor model, but later studies report its morphology cannot be perfectly explained by any available physical models for thermal-composite SNRs \citep{Rho02}. We interpret that the shell-like and the thermal-composite morphology can co-exist in one SNR due to different environments in different directions. In the first 4--5\,kyr, W28 might evolve as a typical shell-like SNR until having a large size of tens of parsecs. After that, it is reasonable that different parts of a middle-aged SNR evolve in different environments. Once in a dense medium, the cloud evaporation and thermal conduction may be significant, leading to centrally brightened X-ray emissions with recombination features, while in other directions with lower density, a continuous evolution in the Sedov-Taylor phase formed the western shell of W28. 

\section{Conclusion}
\label{sec: sum}
We showed the EP-FXT studies on the prototypical thermal-composite SNR W28. The large FOV of FXT maps this large SNR with high efficiency and provides new insights into this remnant. We conclude them as below:

\begin{itemize}
\item  New X-ray shell-like emission feature is discovered to the west of the W28 center and spatially associated with its counterparts in radio and optical wavelengths. It shares similar pressure and foreground absorption with the gas in the W28 center, but has a high electron temperature of $\sim1.5$\,keV and is underionized. These properties, together with the shell orientation, suggest that they are part of the SNR W28.  Adopting this assumption, we estimate a dynamical age of $\sim8$\,kyr, much younger than previous studies, and explosion energy of $\sim(1$--$2)\times10^{51}$\,erg.

\item 

The EP-FXT observation reveals a complex morphology of W28 and its vicinity with a western shell and several clumps apart from its thermal-composite center. The spectral analysis presents the spatial distribution of ionization status of the hot plasma in W28. Recombining plasma with a low electron temperature of $\sim 0.2$\,keV and initial temperature of $\sim0.6$--$0.7$\,keV is spatially coincident with H$\alpha$ emission in the W28 center. The other regions show plasma in ionizing or CIE states and the western shell has a high temperature of $>1$\,keV. Generally, the heavy element abundances of the hot gas are subsolar.

\item
We found a similarity between the initial temperature $t_\mathrm{\rm init}$ of the recombining plasma and the current temperature $T$ of the CIE plasma and a near smooth distribution of $\max(t_\mathrm{\rm init}, T)$ in most regions of W28. This suggests that most area of W28  were generally isothermal with a temperature of $\sim0.7$\,keV before the plasma cooling.

\item
We use a kT-EM diagram of the regions and found a near isobaric distribution. This suggests the western shell and the W28 center may share the same physical origin and the recombination in W28 is probably dominated by thermal conduction instead of adiabatic expansion. We estimate a recombination timescale of $\sim3$\,kyr, which is more likely to arise from the short-range saturated thermal conduction and cloud evaporation instead of long-range conduction.

\item The complex morphology and ionization state distribution of W28 are probably the result of its evolution in an environment with high density gradient. The shell-like and thermal-composite morphology could co-exist in one SNR, as the morphology originates more from nurture than nature. 

\end{itemize}
\begin{acknowledgements}
    We would like to express our sincere gratitude to the referee for their constructive comments, which have greatly improved the paper.
    We thank Fangjun Lu for helpful discussions in SNR science with EP and for requesting the W28 observations. This work is based on the data obtained with Einstein Probe, a space mission supported by the Strategic Priority Program on Space Science of Chinese Academy of Sciences, in collaboration with the European Space Agency, the Max-Planck-Institute for extraterrestrial Physics (Germany), and the Centre National d'Études Spatiales (France). This study is based on observations obtained with XMM-Newton, an ESA science mission with instruments and contributions directly funded by ESA Member States and NASA. This research made use of the data from the Milky Way Imaging Scroll Painting (MWISP) project, which is a multi-line survey in 12CO/13CO/C18O along the northern galactic plane with PMO-13.7m telescope. We are grateful to all the members of the MWISP working group, particularly the staff members at PMO-13.7m telescope, for their long-term support. MWISP was sponsored by National Key R\&D Program of China with grant 2017YFA0402701 and CAS Key Research Program of Frontier Sciences with grant QYZDJ-SSW-SLH047. Y.-H. C. acknowledges supports from the National Natural Science Foundation of China (NSFC) via grant NSFC 123B1021. P. Z.\ thanks the support by NSFC 12273010, the Fundamental Research Funds for the Central Universities No.\ KG202502, the China Manned Space Program with grant nos. CMS-CSST-2025-A14 and CMS-CSST-2025-A18. Y.C.\ acknowledges NSFC under grants 12121003 \& 12573047. L.S. thanks the support from NSFC grant No. 12503030. C. G. acknowledges the foundation from the grant NFSC 12373007 \& 12422302.
\end{acknowledgements}

\bibliographystyle{aa}
\bibliography{aa}

\begin{appendix}
\section{Details of the X-ray observations}\label{app: obs}
The details of the EP-FXT and XMM-Newton observations are listed in Table~\ref{tab: obs}.

\setlength\tabcolsep{1pt}{
\begin{table}[!htbp]
    \centering
    \footnotesize
    \caption{EP-FXT and XMM-Newton observations.}
    \begin{tabular}{cccccccc}
    \hline\hline
        Obs ID & PI & Obs. & R.A. & Decl. & \multicolumn{3}{c}{Exposure$^a$} \\
        ~ & ~ & Date & (J2000) & (J2000) & ~ & (ks) & ~ \\ \hline
        \multicolumn{5}{c}{EP-FXT} & FXTA & ~ & FXTB \\ \hline
        11912598785 & Lu & 24-04-04 & 270.30 & $-23.48$ & 5.9 & ~ & 5.9 \\
        11912599297 & Lu & 24-04-04 & 270.76 & $-23.48$ & 8.9 & ~ & 8.9 \\ \hline
        \multicolumn{5}{c}{XMM/EPIC} & MOS1 & MOS2 & pn \\ \hline
        0135742201 & Warwick & 03-03-19 & 270.07 & $-23.32$ & 5.7 & 6.0 & 3.6 \\
		0135742301 & Warwick & 03-03-20 & 269.89 & $-23.03$ & 2.9 & 3.5 & 1.5 \\
		0135742401 & Warwick & 03-03-20 & 270.43 & $-23.31$ & 5.4 & 4.1 & 0.6 \\
		0135742501 & Warwick & 03-03-20 & 270.25 & $-23.03$ & 2.8 & 2.9 & 1.2 \\
		0145970101$^b$ & Koyama & 02-09-23 & 270.46 & $-23.32$ & 20.5 & 20.9 & 15.0 \\
		0145970401$^b$ & Koyama & 03-10-07 & 270.46 & $-23.32$ & 20.9 & 21.4 & 14.5 \\
		0503850101 & Funk & 08-04-03 & 270.10 & $-24.04$ & 27.9 & 31.5 & 17.9 \\
		0886090501 & Ponti & 23-10-11 & 270.21 & $-23.86$ & 8.6 & 10.1 & 6.6 \\
		0886110201 & Ponti & 24-03-17 & 269.76 & $-23.81$ & 7.3 & 9.5 & 4.5 \\
		0886110301 & Ponti & 23-10-10 & 269.59 & $-23.53$ & 16.1 & 17.8 & 12.1 \\
		0886110501 & Ponti & 23-10-13 & 270.42 & $-23.53$ & 19.8 & 21.1 & 10.1 \\
		0886110601 & Ponti & 23-10-10 & 269.38 & $-23.20$ & 13.4 & 16.3 & 6.1 \\
		0886110701 & Ponti & 24-04-04 & 269.97 & $-23.48$ & 21.3 & 21.2 & 17.0 \\ \hline\hline
        \end{tabular}
        \tablefoot{\tablefoottext{a}{For XMM-Newton the exposure times are flare-filtered.} \\ \tablefoottext{b}{With Thick filter. All the other XMM-Newton observations use Medium filter.}}
    \label{tab: obs}
\end{table}
}

\section{FXT v.s. XMM-Newton: calibration and limitation}
\label{app: cal}

The EP-FXT PV phase observations were performed before the completion of the first in-orbit calibration. Although the raw data have been updated with the latest calibration database, it is still necessary to use reliable and well-calibrated data from other X-ray telescopes for cross-calibration. With high surface brightness, the NE shell (Region~44, in Fig.~\ref{fig: 3bandrgb}) is an appropriate source for calibration in our field of view. Besides, located in the upper part of the detector, or rather further from GX~5$-$1, this region appears to suffer from minimal contamination of the stray light.

\begin{figure}[!htbp]
    \centering
    \includegraphics[width=0.49\textwidth]{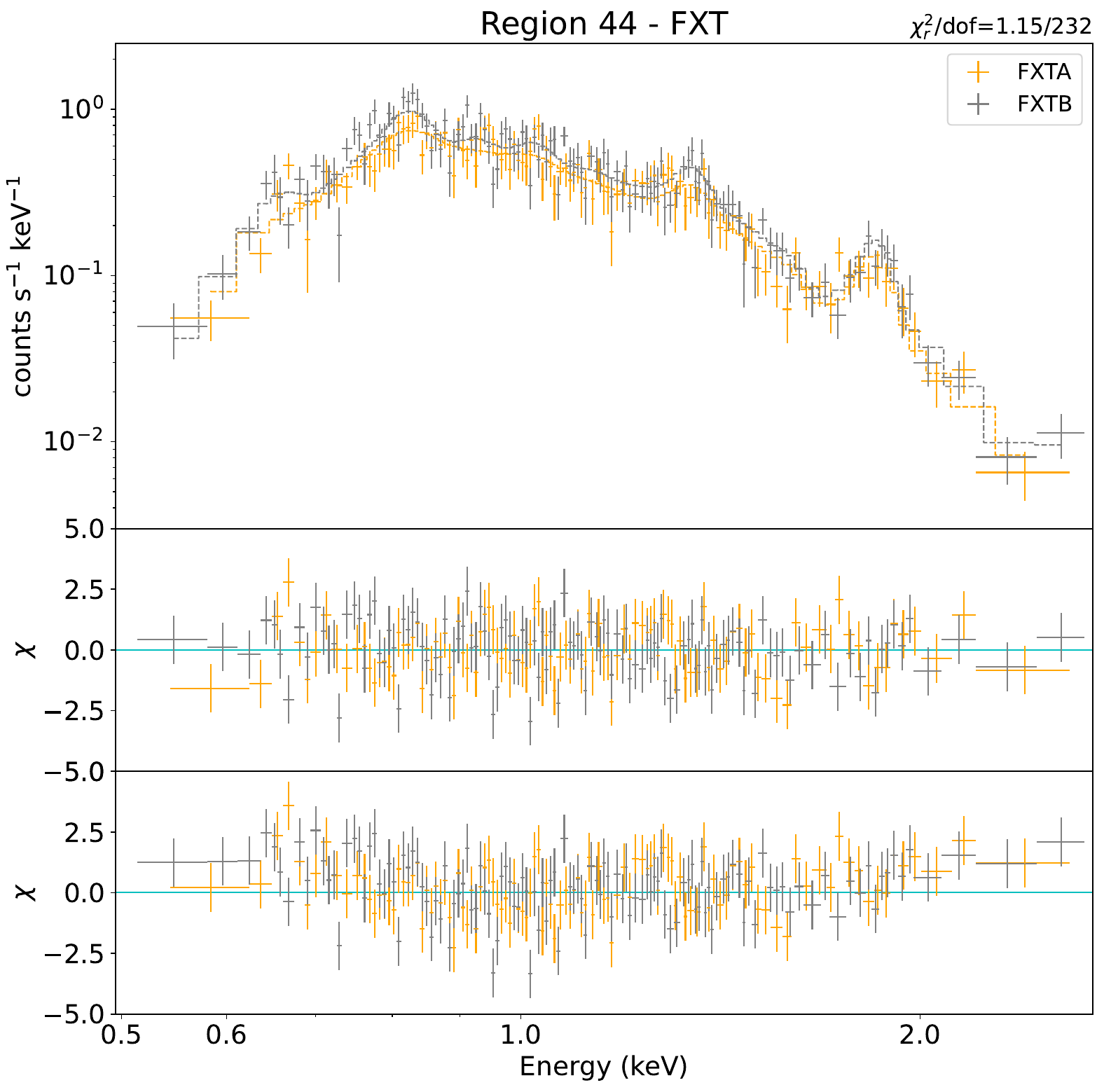}
    \caption{Upper panel: FXTA (marked in orange) and FXTB (marked in grey) spectra of the NE shell (Region 44). Middle panel: residuals of the best-fit model. Lower panel: residuals to compare the FXT spectra with the XMM-Newton model after rescaling the normalizations.}
    \label{fig: fxt44}
\end{figure}

\begin{figure}[!htbp]
    \centering
    \includegraphics[width=0.49\textwidth]{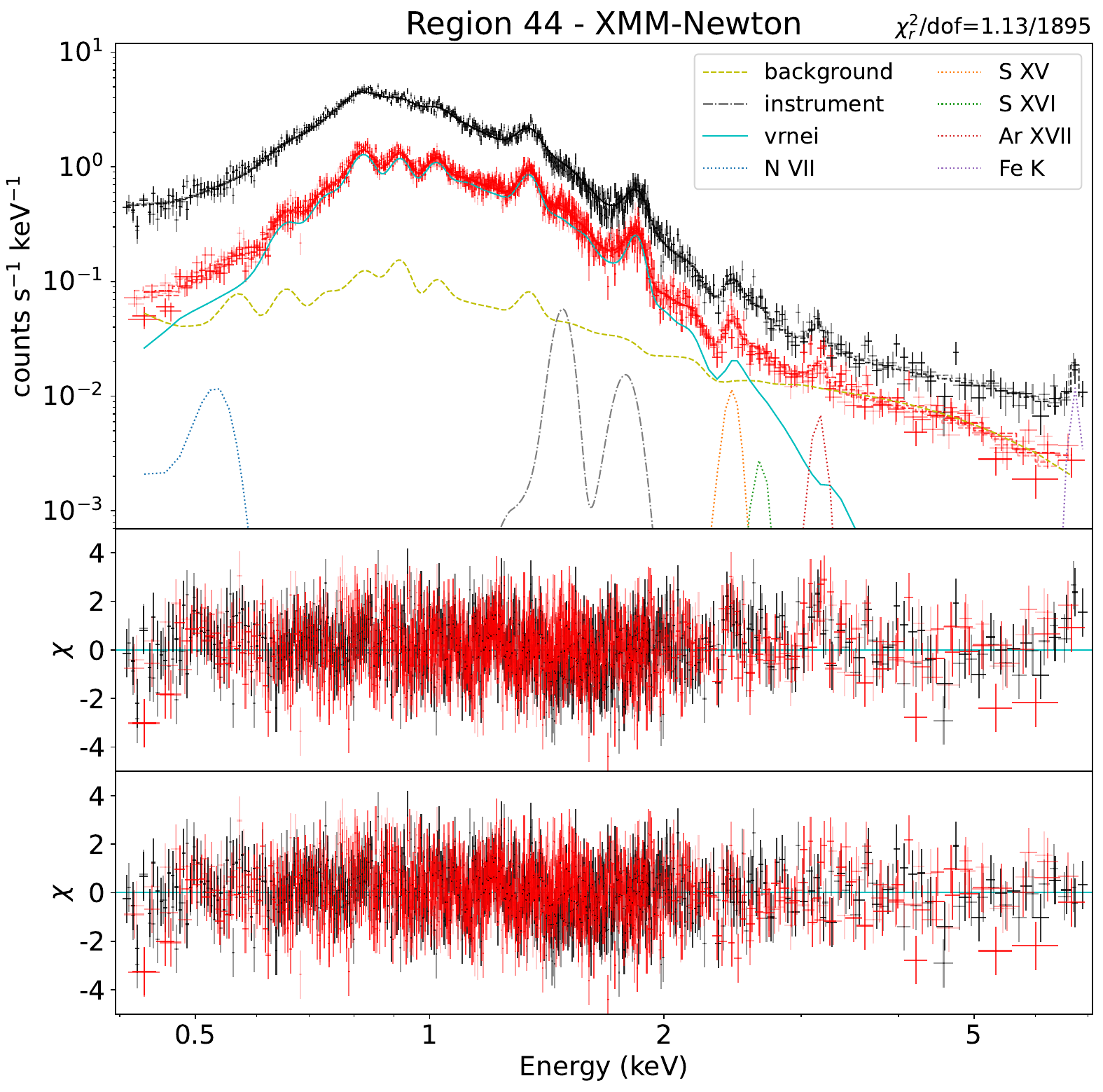}
    \caption{Upper panel: XMM-Newton pn (marked in black) and MOS (marked in red) spectra of the NE shell (Region 44). Dashed lines show the model folded with the response file. Different levels of transparency stand for different observations and different cameras (i.e., MOS 1\&2). The components of one MOS spectra were displayed as well, including the astrophysical background (the yellow dashed line), instrument fluorescence lines (Al and Si, the grey dash-dot line), overionized plasma (the cyan solid line), and extra {\it gaussian} components for emission lines (dotted lines in different colors). Middle panel: residuals of the best-fit model without extra line components. Lower panel: residuals of the best-fit panel.}
    \label{fig: xmmfit}
\end{figure}

We first used a CIE model ({\it vapec}) to fit the FXT spectra in 0.5--2.0\,keV with good statistic quality and minimal influence from the stray light. The best-fit model ($\chi^2_r$/dof=1.35/233, where dof is short for degrees of freedom) gives an absorption of $N_\mathrm{H}$=$7.3\times10^{21}$\,cm$^{-2}$, a temperature of 0.47\,keV, and subsolar Ne, Mg, Si, and Fe. However, a residual appears near 0.9\,keV, which is more evident if all the XMM-Newton and FXT spectra are fitted jointly. Therefore, the CIE model is substituted by an overionized NEI model ({\it vrnei}), effectively improving the fitting statistics to $\chi^2_r$/dof=1.15/232 (see Table~\ref{tab: cal} and Fig.~\ref{fig: fxt44}). The abundance of O, Ne, Mg, and Si seems subsolar while the Fe abundance cannot be constrained well and thus fixed to solar abundance as other metals. The ionization timescale is around $10^{12}$\,s\,cm$^{-3}$, quite close to CIE and probably leading to a poor constraint of the initial temperature.

\setlength\tabcolsep{1pt}
\begin{table}[t!]
    \centering
    \footnotesize
    \caption{Fitting results of Region 44 and 26 for calibration with 1$\sigma$ uncertainties.}
    \begin{tabular}{ccccccc}
    \hline
        Comp. & Para. & Unit & \multicolumn{2}{c}{Region 44} & \multicolumn{2}{c}{Region 26} \\ 
        ~ & ~ & ~ & FXT & XMM & FXT & XMM \\ \hline
{\it TBabs} & $N_\mathrm{H}$ & $10^{21}$\,cm$^{-2}$ & $11.98_{-0.34}^{+0.30}$ & $12.88_{-0.05}^{+0.07}$ & $7.85_{-0.60}^{+0.52}$ & $9.36_{-0.21}^{+0.22}$ \\ \hline
~ & $kT$ & keV & $0.19_{-0.03}^{+0.03}$ & $0.225_{-0.002}^{+0.002}$ & $0.16_{-0.01}^{+0.03}$ & $0.150_{-0.006}^{+0.005}$ \\
~ & $kT_\mathrm{init}$ & keV & $>1.43$ & $1.51_{-0.32}^{+0.40}$ & $1.00_{-0.11}^{+0.09}$ & $0.79_{-0.04}^{+0.04}$ \\
~ & O & Z/Z$_\odot$ & $0.46_{-0.25}^{+0.72}$ & $0.65_{-0.04}^{+0.02}$ & $1.01_{-0.43}^{+0.58}$ & $0.83_{-0.12}^{+0.10}$ \\
~ & Ne & Z/Z$_\odot$ & $0.29_{-0.12}^{+0.09}$ & $0.69_{-0.03}^{+0.02}$ & $0.21_{-0.09}^{+0.13}$ & $0.03_{-0.02}^{+0.02}$ \\
{\it vrnei} & Mg & Z/Z$_\odot$ & $0.43_{-0.14}^{+0.06}$ & $0.68_{-0.01}^{+0.01}$ & $0.34_{-0.11}^{+0.23}$ & $0.17_{-0.03}^{+0.04}$ \\
~ & Si & Z/Z$_\odot$ & $0.52_{-0.10}^{+0.15}$ & $1.24_{-0.10}^{+0.10}$ & $0.44_{-0.17}^{+0.38}$ & $0.31_{-0.05}^{+0.06}$ \\
~ & $\tau$ & $10^{11}$\,s\,cm$^{-3}$ & $10.0_{-3.9}^{+3.1}$ & $12.9_{-0.2}^{+0.3}$ & $1.55_{-0.50}^{+0.56}$ & $0.89_{-0.12}^{+0.14}$ \\
~ & EM & cm$^{-5}$ & $0.55_{-0.12}^{+0.38}$ & $0.29_{-0.02}^{+0.03}$ & $0.049_{-0.023}^{+0.047}$ & $0.077_{-0.012}^{+0.017}$ \\ \hline
{\it gaussian} & $E_\mathrm{c}$ & eV & $\cdots$ & $532_{-3}^{+8}$ & $\cdots$ & $\cdots$ \\
(\ion{N}{VII}?) & $norm$ & $10^{-2}$ & $\cdots$ & $1.0_{-0.3}^{+0.3}$ & $\cdots$ & $\cdots$ \\ \hline
{\it gaussian} & $E_\mathrm{c}$ & keV & $\cdots$ & $2.45_{-0.01}^{+0.01}$ & $\cdots$ & $\cdots$ \\
(\ion{S}{XV}) & $norm$ & $10^{-6}$ & $\cdots$ & $8.2_{-2.2}^{+2.0}$ & $\cdots$ & $\cdots$ \\ \hline
{\it gaussian} & $E_\mathrm{c}$ & keV & $\cdots$ & $2.67_{-0.03}^{+0.02}$ & $\cdots$ & $\cdots$ \\
(\ion{S}{XVI}) & $norm$ & $10^{-6}$ & $\cdots$ & $1.9_{-0.9}^{+0.9}$ & $\cdots$ & $\cdots$ \\ \hline
{\it gaussian} & $E_\mathrm{c}$ & keV & $\cdots$ & $3.16_{-0.02}^{+0.01}$ & $\cdots$ & $\cdots$ \\
(\ion{Ar}{XVII}) & $norm$ & $10^{-6}$ & $\cdots$ & $4.4_{-0.7}^{+0.7}$ & $\cdots$ & $\cdots$ \\ \hline
{\it gaussian} & $E_\mathrm{c}$ & keV & $\cdots$ & $6.75_{-0.06}^{+0.03}$ & $\cdots$ & $\cdots$ \\
(Fe-K) & $norm$ & $10^{-6}$ & $\cdots$ & $4.4_{-1.0}^{+1.0}$ & $\cdots$ & $\cdots$ \\ \hline
$\chi^2_r$/dof & ~ & ~ & 1.15/232 & 1.13/1895 & 1.02/260 & 1.34/618 \\ \hline
    \end{tabular}
    \label{tab: cal}
\end{table}

Then, a similar model was used to fit the XMM-Newton spectra (see Table~\ref{tab: cal} and Fig.~\ref{fig: xmmfit}). For each spectrum, a {\it constant} component is multiplied to reflect its relative intensity, which can be affected by the calibration uncertainties, CCD gaps, or bad pixels in the region. Generally, due to the longer exposure and large effective area, the XMM-Newton data can better constrain the parameters. The major difference in the fitting results is that XMM-Newton gives a slightly higher absorption and metal abundance, while the temperature, ionization timescale, and emission measure are consistent within the 1$\sigma$ uncertainty.

The fitted parameters obtained through the XMM-Newton data were then compared to the FXT spectra. The average intensity of FXT and XMM-Newton data is consistent within 10\%. We found that the XMM-Newton model fit the FXT spectra best in the 0.8--2.0\,keV band, which covers the line emissions of Fe, Ne, Mg, and Si. However, in lower and higher energy bands with low count rates, FXT data points seem to be higher than the XMM-Newton model.

Noticeably, one single {\it vrnei} model cannot explain the entire XMM-Newton spectra. Evident residuals are present near 0.5, 2.4, 2.6, 3.1, and 6.7\,keV (Fig.~\ref{fig: xmmfit}), probably corresponding to the \ion{N}{VII}, \ion{S}{XV}, \ion{S}{XVI}, \ion{Ar}{XVII}, and \ion{Fe}{K} lines or related RRC, respectively. For instance, the likely S and Ar line features are expected to originate from an extra higher-temperature component to explain the excess above 2\,keV in the FXT spectra, precluding very high non-thermal emissions scenario. Therefore, we added 5 extra {\it gaussian} components to the {\it vrnei} model to improve the fitting.

We also compared the spectra between two missions of Region 26, where the X-ray surface brightness peaks in the W28 center. With a lower exposure and far from the optical axis of XMM-Newton, we simply listed the fitting results in Table~\ref{tab: cal}. The results are similar to those of Region 44. FXT gives a lower column density than XMM-Newton, which may correspond to higher FXT data points of FXT below $\sim0.8$\,keV as mentioned before and plotted in Fig.~\ref{fig: fxt44}. The abundances of heavy elements are generally subsolar in spite of some differences in some specific metals like Ne. The electron temperature, emission measure, and ionization timescale are generally consistent within $\sim1\sigma$ uncertainties.

\section{Fitting results with solar abundance.}
\label{app: fitsolar}
\begin{figure}[ht]
    \includegraphics[width=0.495\textwidth]{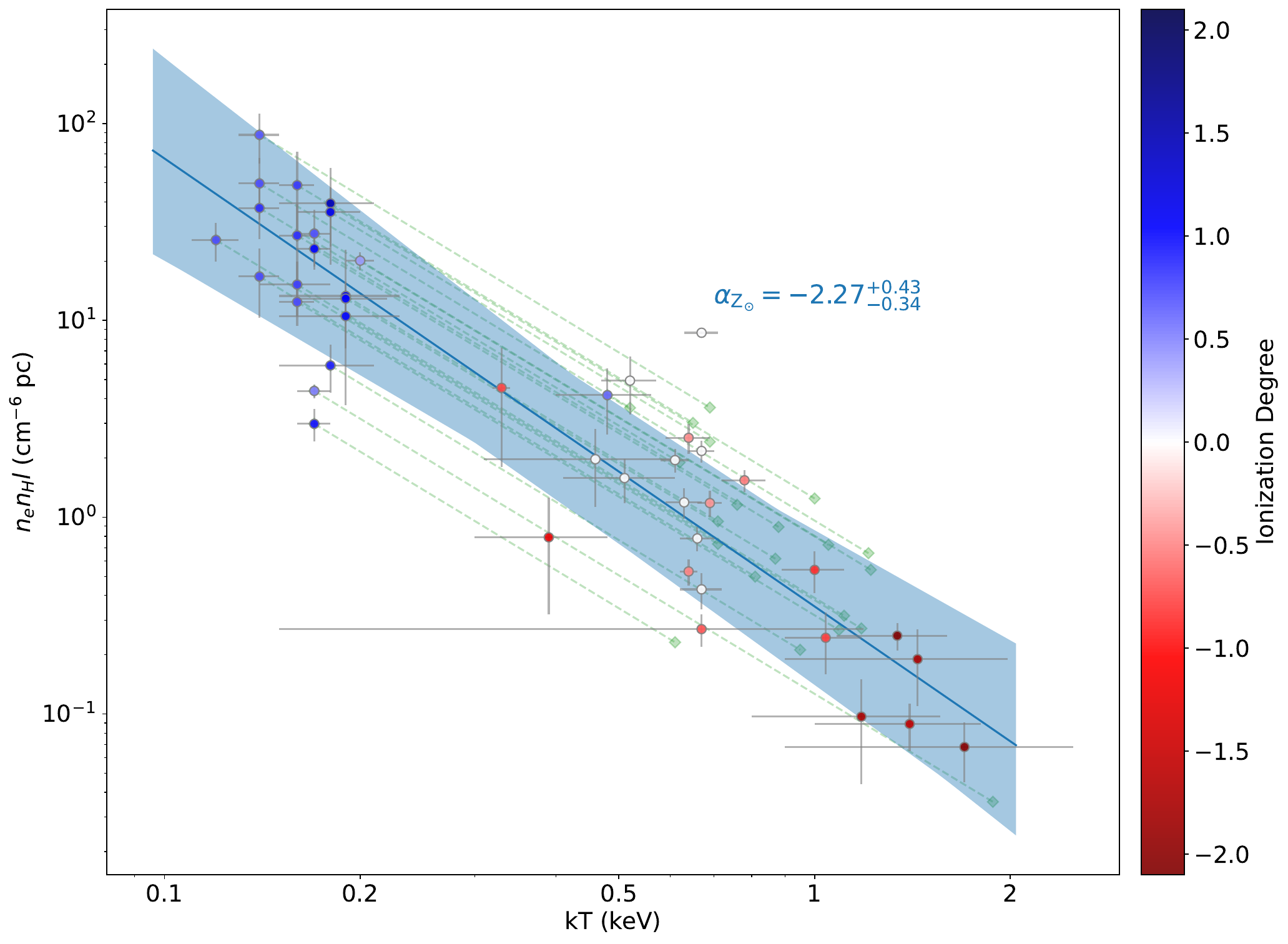}
    \caption{The temperature-EM the same as Fig.~\ref{fig: phase3}, but the data points are based on the spectral analysis with default solar abundances. The uncertainty of index and the confidence belt correspond to $1\sigma$ confidence level due to a larger data dispersion here.}
    \label{fig: phase1}
\end{figure}

\begin{figure}[ht]
    \centering
    \includegraphics[width=\textwidth]{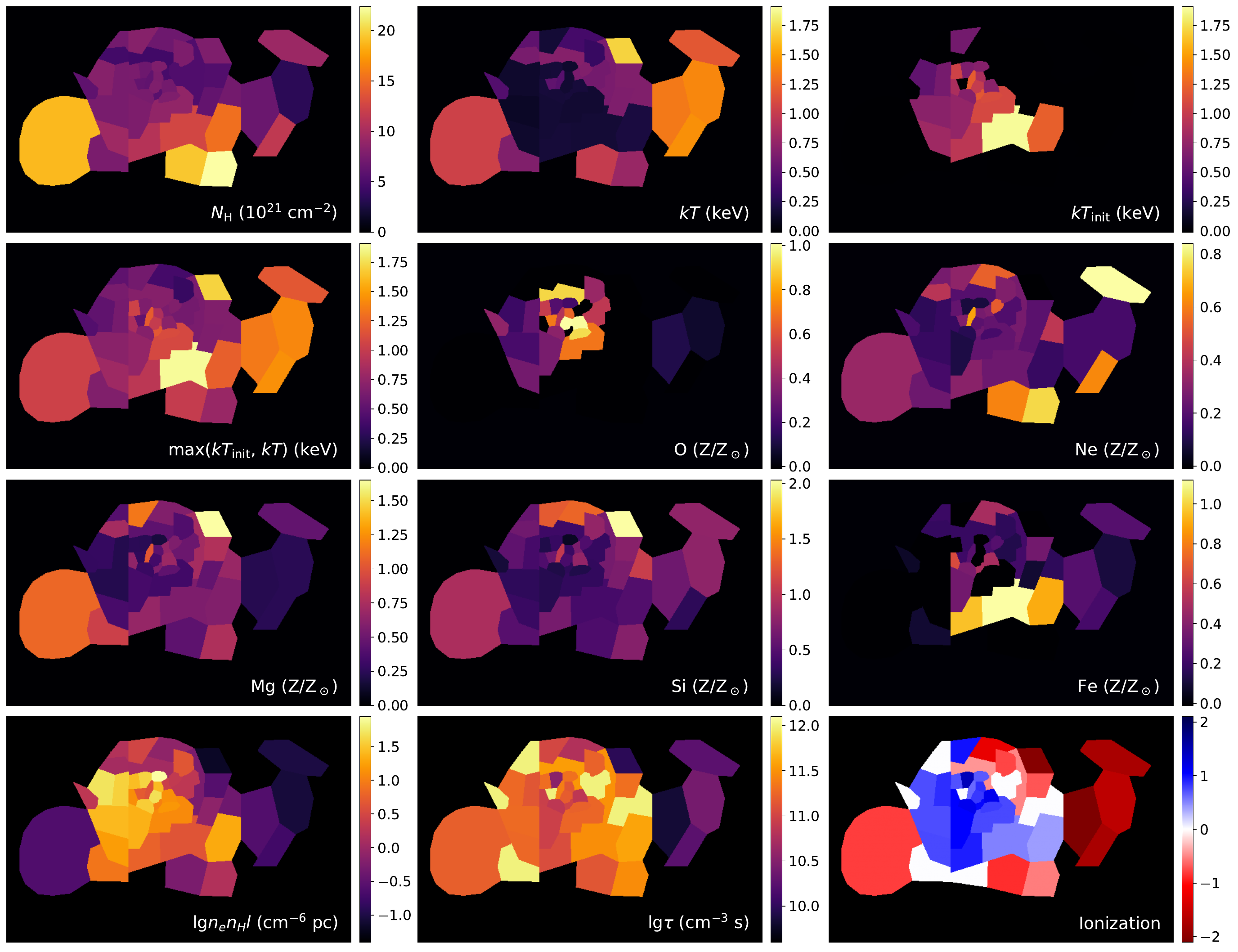}
    \caption{Maps of fitting result (Z=Z$_\odot$). Fixed parameters are set to blank here.}
    \label{fig: resultmap}
\end{figure}

Here shows the fitting parameters with solar abundance in default, including Table~\ref{tab: fit}, the temperature-EM map Fig.~\ref{fig: phase1}, and the parameter map Fig.~\ref{fig: resultmap}.

\onecolumn
\clearpage

\setlength{\LTcapwidth}{\textwidth}
\setlength\tabcolsep{3pt}
\small
\begin{longtable}{ccccccccccccc}
    \caption{Fitting results (Z=Z$_\odot$) with 1$\sigma$ error.}
    \label{tab: fit} \\
    \hline
        Region & $N_\mathrm{H}$ & $kT$ & $kT_\mathrm{init}$ & O & Ne & Mg & Si & S & Fe & $\tau^a$ & $n_\mathrm{e}n_\mathrm{H}l$ & $\chi^2_r$/dof \\
        ~ & ($10^{21}$\,cm$^{-2}$) & (keV) & (keV) & (Z/Z$_\odot$) & (Z/Z$_\odot$) & (Z/Z$_\odot$) & (Z/Z$_\odot$) & (Z/Z$_\odot$) & (Z/Z$_\odot$) & ($10^{10}$\,s\,cm$^{-3}$) & (cm$^{-6}$\,pc) & ~ \\ 
        \hline
    \endfirsthead
    \caption{(continued)}\\ 
    \hline
        Region & $N_\mathrm{H}$ & $kT$ & $kT_\mathrm{init}$ & O & Ne & Mg & Si & S & Fe & $\tau^a$ & $n_\mathrm{e}n_\mathrm{H}l$ & $\chi^2_r$/dof \\
        ~ & ($10^{21}$\,cm$^{-2}$) & (keV) & (keV) & (Z/Z$_\odot$) & (Z/Z$_\odot$) & (Z/Z$_\odot$) & (Z/Z$_\odot$) & (Z/Z$_\odot$) & (Z/Z$_\odot$) & ($10^{10}$\,s\,cm$^{-3}$) & (cm$^{-6}$\,pc) & ~ \\ 
        \hline
    \endhead
    \hline
    \endfoot
    \hline
        \multicolumn{13}{l}{{\bf Notes:} $^a$Ionization timescale $\tau=n_\mathrm{e}t$. No number means CIE. $^b$ Not mapped in Fig.~\ref{fig: resultmap}.} \\
    \endlastfoot
1 & $9.4_{-1.5}^{+1.6}$ & $1.18_{-0.38}^{+1.85}$ & $\cdots$ & $\cdots$ & $0.84_{-0.12}^{+0.26}$ & $0.46_{-0.23}^{+0.33}$ & $0.80_{-0.32}^{+0.46}$ & $\cdots$ & $0.27_{-0.18}^{+0.28}$ & $1.8_{-0.9}^{+18.6}$ & $0.097_{-0.053}^{+0.193}$ & 1.07/54 \\
2 & $3.4_{-1.2}^{+1.4}$ & $1.4_{-0.4}^{+1.6}$ & $\cdots$ & $<0.07$ & $<0.17$ & $0.24_{-0.09}^{+0.11}$ & $0.79_{-0.17}^{+0.21}$ & $\cdots$ & $0.11_{-0.05}^{+0.05}$ & $2.6_{-1.2}^{+3.1}$ & $0.089_{-0.024}^{+0.055}$ & 1.10/96 \\
3 & $11.5_{-1.4}^{+1.3}$ & $1.44_{-0.54}^{+1.32}$ & $\cdots$ & $\cdots$ & $0.62_{-0.13}^{+0.14}$ & $0.24_{-0.12}^{+0.17}$ & $0.31_{-0.13}^{+0.15}$ & $\cdots$ & $<0.23$ & $1.78_{-0.52}^{+1.14}$ & $0.19_{-0.08}^{+0.17}$ & 0.93/111 \\
4 & $6.67_{-0.48}^{+0.61}$ & $1.34_{-0.26}^{+0.15}$ & $\cdots$ & $0.12_{-0.03}^{+0.03}$ & $0.17_{-0.03}^{+0.03}$ & $0.23_{-0.05}^{+0.06}$ & $0.63_{-0.15}^{+0.25}$ & $\cdots$ & $0.27_{-0.09}^{+0.20}$ & $0.69_{-0.17}^{+0.37}$ & $0.25_{-0.04}^{+0.09}$ & 1.04/237 \\
5 & $7.8_{-1.3}^{+1.4}$ & $1.7_{-0.8}^{+1.4}$ & $\cdots$ & $\cdots$ & $\cdots$ & $1.65_{-0.32}^{+0.40}$ & $2.03_{-0.56}^{+0.89}$ & $\cdots$ & $\cdots$ & $0.94_{-0.22}^{+0.37}$ & $0.068_{-0.023}^{+0.050}$ & 0.91/58 \\
6 & $5.4_{-1.2}^{+1.4}$ & $0.67_{-0.52}^{+0.52}$ & $\cdots$ & $\cdots$ & $0.21_{-0.12}^{+0.14}$ & $0.78_{-0.21}^{+0.28}$ & $0.76_{-0.26}^{+0.33}$ & $\cdots$ & $0.35_{-0.10}^{+0.14}$ & $20.3_{-7.1}^{+13.9}$ & $0.27_{-0.05}^{+0.07}$ & 1.01/106 \\
7 & $7.2_{-1.9}^{+1.8}$ & $0.67_{-0.05}^{+0.07}$ & $\cdots$ & $\cdots$ & $<0.42$ & $0.68_{-0.24}^{+0.36}$ & $1.08_{-0.25}^{+0.33}$ & $\cdots$ & $0.17_{-0.09}^{+0.16}$ & $\cdots$ & $0.43_{-0.09}^{+0.10}$ & 0.74/74 \\
8 & $6.34_{-0.64}^{+0.69}$ & $0.66_{-0.04}^{+0.04}$ & $\cdots$ & $\cdots$ & $0.21_{-0.13}^{+0.14}$ & $0.46_{-0.12}^{+0.15}$ & $0.69_{-0.13}^{+0.14}$ & $\cdots$ & $0.09_{-0.02}^{+0.03}$ & $\cdots$ & $0.78_{-0.11}^{+0.14}$ & 1.25/173 \\
8$^b$ & $10.90_{-0.85}^{+0.92}$ & $0.15_{-0.02}^{+0.02}$ & $1.06_{-0.20}^{+0.25}$ & $0.23_{-0.09}^{+0.15}$ & $0.14_{-0.05}^{+0.09}$ & $0.23_{-0.06}^{+0.10}$ & $0.26_{-0.07}^{+0.11}$ & $\cdots$ & $2.9_{-1.9}^{+3.8}$ & $40.2_{-12.4}^{+10.8}$ & $21.3_{-7.9}^{+9.9}$ & 1.21/170 \\
9 & $15.28_{-0.05}^{+0.06}$ & $0.20_{-0.01}^{+0.01}$ & $1.22_{-0.10}^{+0.11}$ & $\cdots$ & $0.14_{-0.05}^{+0.05}$ & $0.59_{-0.05}^{+0.05}$ & $0.49_{-0.06}^{+0.06}$ & $\cdots$ & $0.91_{-0.31}^{+0.48}$ & $39.9_{-4.6}^{+4.7}$ & $20.1_{-2.1}^{+2.6}$ & 1.22/467 \\
10 & $22.37_{-0.89}^{+0.96}$ & $0.78_{-0.06}^{+0.06}$ & $\cdots$ & $\cdots$ & $0.76_{-0.48}^{+0.80}$ & $0.77_{-0.15}^{+0.18}$ & $0.74_{-0.11}^{+0.13}$ & $\cdots$ & $\cdots$ & $29.5_{-10.3}^{+17.8}$ & $1.54_{-0.20}^{+0.23}$ & 0.99/142 \\
11 & $5.32_{-0.68}^{+0.68}$ & $0.64_{-0.02}^{+0.04}$ & $\cdots$ & $\cdots$ & $0.27_{-0.15}^{+0.22}$ & $0.73_{-0.18}^{+0.21}$ & $0.66_{-0.16}^{+0.17}$ & $\cdots$ & $0.22_{-0.04}^{+0.05}$ & $>31.6$ & $0.53_{-0.08}^{+0.10}$ & 1.03/166 \\
12 & $12.67_{-0.74}^{+0.72}$ & $0.17_{-0.01}^{+0.02}$ & $1.88_{-0.42}^{+0.64}$ & $\cdots$ & $0.25_{-0.07}^{+0.08}$ & $0.57_{-0.08}^{+0.08}$ & $0.43_{-0.07}^{+0.09}$ & $0.33_{-0.12}^{+0.11}$ & $1.74_{-0.86}^{+0.85}$ & $30.2_{-3.9}^{+3.1}$ & $4.38_{-0.35}^{+0.71}$ & 1.01/421 \\
13 & $19.44_{-0.69}^{+1.38}$ & $1.00_{-0.11}^{+0.21}$ & $\cdots$ & $\cdots$ & $<0.61$ & $0.44_{-0.15}^{+0.17}$ & $0.46_{-0.11}^{+0.12}$ & $\cdots$ & $\cdots$ & $13.8_{-6.6}^{+13.0}$ & $0.54_{-0.13}^{+0.12}$ & 1.00/127 \\
14 & $6.9_{-2.5}^{+1.9}$ & $0.39_{-0.09}^{+0.24}$ & $\cdots$ & $\cdots$ & $0.54_{-0.17}^{+0.27}$ & $0.59_{-0.17}^{+0.32}$ & $1.34_{-0.81}^{+2.28}$ & $\cdots$ & $<0.5$ & $>6.1$ & $0.79_{-0.47}^{+0.53}$ & 0.93/75 \\
15 & $7.73_{-2.61}^{+0.98}$ & $0.33_{-0.01}^{+0.18}$ & $\cdots$ & $0.41_{-0.23}^{+0.64}$ & $0.29_{-0.12}^{+0.27}$ & $0.39_{-0.13}^{+0.21}$ & $0.81_{-0.40}^{+0.38}$ & $\cdots$ & $0.17_{-0.07}^{+0.13}$ & $>17.0$ & $4.54_{-2.74}^{+1.70}$ & 1.02/217 \\
16 & $5.29_{-0.52}^{+0.55}$ & $0.67_{-0.03}^{+0.03}$ & $\cdots$ & $\cdots$ & $0.53_{-0.19}^{+0.21}$ & $0.58_{-0.16}^{+0.19}$ & $0.47_{-0.13}^{+0.15}$ & $\cdots$ & $0.27_{-0.04}^{+0.05}$ & $\cdots$ & $2.17_{-0.28}^{+0.31}$ & 1.09/157 \\
17 & $5.29_{-0.55}^{+0.59}$ & $0.61_{-0.03}^{+0.02}$ & $\cdots$ & $0.51_{-0.29}^{+0.46}$ & $0.19_{-0.10}^{+0.13}$ & $0.50_{-0.10}^{+0.13}$ & $0.35_{-0.06}^{+0.08}$ & $\cdots$ & $0.14_{-0.03}^{+0.05}$ & $\cdots$ & $1.95_{-0.27}^{+0.32}$ & 1.05/345 \\
18 & $8.89_{-0.28}^{+0.26}$ & $0.16_{-0.01}^{+0.01}$ & $1.09_{-0.08}^{+0.10}$ & $0.70_{-0.13}^{+0.19}$ & $0.23_{-0.05}^{+0.07}$ & $0.33_{-0.07}^{+0.10}$ & $0.34_{-0.07}^{+0.11}$ & $0.22_{-0.14}^{+0.16}$ & $\cdots$ & $17.6_{-1.5}^{+2.8}$ & $12.4_{-3.0}^{+3.8}$ & 1.11/548 \\
18$^b$ & $9.27_{-0.36}^{+0.34}$ & $0.15_{-0.01}^{+0.01}$ & $2.01_{-0.50}^{+1.00}$ & $\cdots$ & $0.39_{-0.06}^{+0.06}$ & $0.45_{-0.05}^{+0.06}$ & $0.38_{-0.05}^{+0.06}$ & $0.17_{-0.09}^{+0.14}$ & $2.48_{-0.43}^{+0.56}$ & $30.1_{-3.2}^{+2.8}$ & $9.72_{-1.05}^{+1.13}$ & 1.11/548 \\
19 & $8.23_{-1.06}^{+0.98}$ & $0.17_{-0.01}^{+0.02}$ & $0.61_{-0.10}^{+0.12}$ & $\cdots$ & $0.32_{-0.18}^{+0.70}$ & $1.16_{-0.34}^{+0.43}$ & $1.27_{-0.28}^{+11.28}$ & $\cdots$ & $\cdots$ & $<10.5$ & $2.98_{-0.56}^{+0.62}$ & 0.99/80 \\
20 & $4.52_{-0.58}^{+0.59}$ & $0.69_{-0.03}^{+0.03}$ & $\cdots$ & $0.91_{-0.49}^{+0.83}$ & $<0.2$ & $0.44_{-0.11}^{+0.14}$ & $0.39_{-0.09}^{+0.11}$ & $\cdots$ & $0.19_{-0.05}^{+0.08}$ & $36.2_{-13.0}^{+18.9}$ & $1.18_{-0.18}^{+0.18}$ & 1.02/276 \\
21$^b$ & $3.79_{-0.65}^{+0.70}$ & $0.61_{-0.05}^{+0.05}$ & $\cdots$ & $0.33_{-0.23}^{+0.38}$ & $0.12_{-0.07}^{+0.14}$ & $0.31_{-0.08}^{+0.11}$ & $0.30_{-0.08}^{+0.10}$ & $\cdots$ & $0.10_{-0.02}^{+0.03}$ & $>27.2$ & $3.59_{-0.66}^{+0.89}$ & 1.15/199 \\
21 & $7.84_{-0.41}^{+0.40}$ & $0.14_{-0.01}^{+0.01}$ & $0.69_{-0.06}^{+0.07}$ & $0.20_{-0.06}^{+0.09}$ & $0.08_{-0.02}^{+0.03}$ & $0.18_{-0.04}^{+0.06}$ & $0.13_{-0.06}^{+0.08}$ & $\cdots$ & $\cdots$ & $20.6_{-2.6}^{+2.8}$ & $87.6_{-25.0}^{+33.8}$ & 1.08/199 \\
22 & $4.85_{-0.70}^{+0.90}$ & $0.64_{-0.05}^{+0.05}$ & $\cdots$ & $\cdots$ & $0.29_{-0.10}^{+0.14}$ & $0.66_{-0.15}^{+0.19}$ & $0.69_{-0.19}^{+0.23}$ & $\cdots$ & $0.24_{-0.04}^{+0.06}$ & $34.0_{-10.0}^{+19.0}$ & $2.53_{-0.44}^{+0.58}$ & 1.08/119 \\
23 & $7.18_{-0.92}^{+1.08}$ & $0.19_{-0.04}^{+0.07}$ & $0.71_{-0.05}^{+0.08}$ & $1.00_{-0.48}^{+2.75}$ & $0.23_{-0.13}^{+0.65}$ & $0.69_{-0.37}^{+1.57}$ & $0.81_{-0.46}^{+1.87}$ & $\cdots$ & $0.50_{-0.27}^{+0.11}$ & $<7.32$ & $13.3_{-9.6}^{+18.9}$ & 1.13/258 \\
24 & $7.80_{-0.48}^{+0.53}$ & $0.16_{-0.02}^{+0.01}$ & $1.11_{-0.19}^{+0.12}$ & $0.92_{-0.45}^{+0.29}$ & $0.17_{-0.11}^{+0.05}$ & $0.45_{-0.19}^{+0.19}$ & $0.26_{-0.13}^{+0.16}$ & $\cdots$ & $\cdots$ & $16.3_{-3.4}^{+3.9}$ & $15.2_{-4.7}^{+20.8}$ & 1.07/238 \\
24$^b$ & $8.20_{-0.51}^{+0.52}$ & $0.15_{-0.01}^{+0.01}$ & $1.36_{-0.19}^{+0.22}$ & $\cdots$ & $0.21_{-0.07}^{+0.08}$ & $0.48_{-0.08}^{+0.09}$ & $0.23_{-0.08}^{+0.08}$ & $\cdots$ & $2.14_{-0.77}^{+1.09}$ & $25.7_{-6.2}^{+5.4}$ & $16.1_{-2.8}^{+3.4}$ & 1.07/238 \\
25$^b$ & $7.58_{-0.72}^{+0.61}$ & $0.17_{-0.01}^{+0.02}$ & $0.92_{-0.13}^{+0.17}$ & $\cdots$ & $0.41_{-0.10}^{+0.12}$ & $0.62_{-0.15}^{+0.23}$ & $0.52_{-0.16}^{+0.20}$ & $\cdots$ & $1.36_{-0.58}^{+1.17}$ & $13.8_{-5.5}^{+7.5}$ & $18.1_{-3.3}^{+3.4}$ & 1.04/168 \\
25 & $7.61_{-0.57}^{+0.49}$ & $0.16_{-0.01}^{+0.02}$ & $0.88_{-0.09}^{+0.10}$ & $0.66_{-0.23}^{+0.41}$ & $0.27_{-0.10}^{+0.19}$ & $0.43_{-0.16}^{+0.38}$ & $0.37_{-0.14}^{+0.32}$ & $\cdots$ & $\cdots$ & $13.6_{-4.4}^{+3.6}$ & $27.0_{-11.8}^{+16.5}$ & 1.04/168 \\
26 & $7.85_{-0.60}^{+0.52}$ & $0.16_{-0.01}^{+0.03}$ & $1.00_{-0.11}^{+0.09}$ & $1.01_{-0.43}^{+0.58}$ & $0.21_{-0.09}^{+0.13}$ & $0.34_{-0.11}^{+0.23}$ & $0.44_{-0.17}^{+0.38}$ & $\cdots$ & $\cdots$ & $15.5_{-5.0}^{+5.6}$ & $48.6_{-23.4}^{+47.4}$ & 1.02/260 \\
26$^b$ & $8.20_{-0.47}^{+0.48}$ & $0.14_{-0.01}^{+0.01}$ & $1.21_{-0.22}^{+0.20}$ & $\cdots$ & $0.24_{-0.06}^{+0.07}$ & $0.33_{-0.07}^{+0.07}$ & $0.38_{-0.08}^{+0.08}$ & $\cdots$ & $3.34_{-1.84}^{+2.07}$ & $26.9_{-10.5}^{+6.5}$ & $31.3_{-5.0}^{+5.6}$ & 1.02/260 \\
27 & $5.71_{-0.92}^{+1.01}$ & $0.19_{-0.04}^{+0.09}$ & $1.18_{-0.16}^{+0.94}$ & $\cdots$ & $0.25_{-0.16}^{+0.22}$ & $0.58_{-0.19}^{+0.49}$ & $0.31_{-0.20}^{+0.27}$ & $\cdots$ & $0.51_{-0.21}^{+0.59}$ & $9.55_{-4.18}^{+13.50}$ & $10.5_{-3.3}^{+4.9}$ & 0.98/95 \\
28 & $6.85_{-0.72}^{+0.70}$ & $0.18_{-0.03}^{+0.07}$ & $0.65_{-0.05}^{+0.07}$ & $0.24_{-0.11}^{+0.23}$ & $0.08_{-0.03}^{+0.09}$ & $0.24_{-0.08}^{+0.20}$ & $0.36_{-0.13}^{+0.32}$ & $\cdots$ & $0.19_{-0.11}^{+0.21}$ & $3.50_{-2.97}^{+5.58}$ & $39.3_{-20.1}^{+25.3}$ & 1.10/233 \\
29 & $6.18_{-0.79}^{+0.82}$ & $0.48_{-0.08}^{+0.05}$ & $1.21_{-0.25}^{+2.15}$ & $0.48_{-0.37}^{+0.76}$ & $0.67_{-0.35}^{+0.74}$ & $1.03_{-0.30}^{+0.68}$ & $1.05_{-0.29}^{+0.53}$ & $\cdots$ & $0.28_{-0.10}^{+0.22}$ & $24.5_{-10.8}^{+20.0}$ & $4.17_{-1.54}^{+3.02}$ & 1.19/161 \\
30 & $7.51_{-0.47}^{+0.47}$ & $0.18_{-0.02}^{+0.02}$ & $0.69_{-0.05}^{+0.08}$ & $0.38_{-0.12}^{+0.21}$ & $0.11_{-0.04}^{+0.06}$ & $0.30_{-0.08}^{+0.13}$ & $0.25_{-0.09}^{+0.13}$ & $\cdots$ & $0.37_{-0.15}^{+0.30}$ & $7.36_{-3.08}^{+4.22}$ & $35.5_{-11.8}^{+16.0}$ & 1.00/263 \\
31 & $7.57_{-0.44}^{+0.41}$ & $0.17_{-0.01}^{+0.02}$ & $1.05_{-0.13}^{+0.14}$ & $0.53_{-0.14}^{+0.19}$ & $0.23_{-0.08}^{+0.13}$ & $0.25_{-0.08}^{+0.11}$ & $0.26_{-0.09}^{+0.12}$ & $\cdots$ & $\cdots$ & $19.5_{-3.0}^{+5.4}$ & $27.6_{-8.8}^{+12.1}$ & 1.00/207 \\
32 & $6.36_{-0.84}^{+0.88}$ & $0.52_{-0.05}^{+0.04}$ & $\cdots$ & $0.71_{-0.49}^{+1.16}$ & $0.29_{-0.16}^{+0.33}$ & $0.26_{-0.11}^{+0.19}$ & $0.63_{-0.16}^{+0.26}$ & $\cdots$ & $0.21_{-0.08}^{+0.16}$ & $\cdots$ & $4.94_{-1.63}^{+2.28}$ & 1.09/166 \\
33 & $7.49_{-0.80}^{+0.66}$ & $0.19_{-0.03}^{+0.04}$ & $0.87_{-0.09}^{+0.12}$ & $\cdots$ & $0.14_{-0.12}^{+0.11}$ & $0.65_{-0.18}^{+0.24}$ & $0.47_{-0.17}^{+0.25}$ & $\cdots$ & $0.65_{-0.27}^{+0.56}$ & $9.01_{-4.06}^{+5.54}$ & $12.9_{-2.8}^{+3.1}$ & 1.12/155 \\
34 & $7.75_{-0.41}^{+0.40}$ & $0.17_{-0.01}^{+0.01}$ & $0.76_{-0.04}^{+0.06}$ & $0.33_{-0.07}^{+0.09}$ & $0.09_{-0.02}^{+0.03}$ & $0.35_{-0.07}^{+0.09}$ & $0.28_{-0.06}^{+0.09}$ & $0.47_{-0.25}^{+0.33}$ & $0.35_{-0.12}^{+0.21}$ & $9.14_{-1.59}^{+4.22}$ & $23.1_{-5.0}^{+5.9}$ & 0.98/504 \\
35 & $10.9_{-1.0}^{+1.1}$ & $0.18_{-0.03}^{+0.06}$ & $0.95_{-0.10}^{+0.27}$ & $\cdots$ & $0.31_{-0.10}^{+0.11}$ & $0.67_{-0.16}^{+0.27}$ & $0.67_{-0.20}^{+0.31}$ & $\cdots$ & $0.96_{-0.55}^{+0.97}$ & $11.7_{-5.1}^{+11.7}$ & $5.9_{-1.6}^{+1.6}$ & 0.93/283 \\
37 & $7.2_{-3.1}^{+2.4}$ & $0.51_{-0.10}^{+0.10}$ & $\cdots$ & $\cdots$ & $0.27_{-0.21}^{+0.41}$ & $0.41_{-0.22}^{+0.40}$ & $0.59_{-0.30}^{+0.41}$ & $\cdots$ & $0.20_{-0.10}^{+0.30}$ & $\cdots$ & $1.58_{-0.40}^{+1.16}$ & 1.09/33 \\
38 & $4.60_{-0.63}^{+0.67}$ & $0.63_{-0.04}^{+0.04}$ & $\cdots$ & $\cdots$ & $0.41_{-0.18}^{+0.21}$ & $0.73_{-0.20}^{+0.26}$ & $0.64_{-0.19}^{+0.22}$ & $\cdots$ & $0.18_{-0.03}^{+0.04}$ & $\cdots$ & $1.19_{-0.21}^{+0.27}$ & 1.15/106 \\
39 & $8.28_{-0.45}^{+0.47}$ & $0.14_{-0.01}^{+0.01}$ & $0.52_{-0.04}^{+0.04}$ & $0.18_{-0.06}^{+0.09}$ & $0.12_{-0.03}^{+0.05}$ & $0.29_{-0.07}^{+0.11}$ & $0.55_{-0.18}^{+0.30}$ & $\cdots$ & $\cdots$ & $18.6_{-2.8}^{+2.89}$ & $49.6_{-17.1}^{+26.4}$ & 1.20/230 \\
40 & $7.44_{-0.39}^{+0.40}$ & $0.14_{-0.01}^{+0.01}$ & $0.62_{-0.05}^{+0.05}$ & $0.28_{-0.09}^{+0.11}$ & $0.14_{-0.03}^{+0.04}$ & $0.22_{-0.05}^{+0.07}$ & $0.43_{-0.11}^{+0.14}$ & $\cdots$ & $\cdots$ & $13.6_{-2.3}^{+2.1}$ & $37.2_{-11.4}^{+18.0}$ & 1.14/244 \\
41$^b$ & $3.84_{-0.26}^{+0.28}$ & $0.69_{-0.02}^{+0.02}$ & $\cdots$ & $0.79_{-0.25}^{+0.33}$ & $0.32_{-0.12}^{+0.14}$ & $1.02_{-0.17}^{+0.20}$ & $0.72_{-0.11}^{+0.13}$ & $\cdots$ & $0.16_{-0.03}^{+0.03}$ & $\cdots$ & $0.87_{-0.10}^{+0.12}$ & 1.15/300 \\
41 & $7.44_{-0.31}^{+0.32}$ & $0.12_{-0.01}^{+0.01}$ & $0.71_{-0.04}^{+0.04}$ & $0.21_{-0.05}^{+0.06}$ & $0.14_{-0.03}^{+0.03}$ & $0.22_{-0.04}^{+0.06}$ & $0.32_{-0.07}^{+0.09}$ & $\cdots$ & $\cdots$ & $18.7_{-1.9}^{+1.9}$ & $25.6_{-5.7}^{+7.9}$ & 1.08/299 \\
42 & $9.53_{-0.58}^{+0.58}$ & $0.14_{-0.01}^{+0.01}$ & $0.81_{-0.08}^{+0.11}$ & $0.32_{-0.11}^{+0.20}$ & $0.17_{-0.06}^{+0.10}$ & $0.36_{-0.10}^{+0.19}$ & $0.37_{-0.10}^{+0.20}$ & $\cdots$ & $\cdots$ & $18.2_{-4.7}^{+4.4}$ & $16.7_{-6.4}^{+8.4}$ & 1.33/170 \\
43 & $7.6_{-1.2}^{+1.5}$ & $0.67_{-0.04}^{+0.05}$ & $\cdots$ & $\cdots$ & $<0.25$ & $0.90_{-0.22}^{+0.27}$ & $0.51_{-0.13}^{+0.15}$ & $\cdots$ & $0.09_{-0.05}^{+0.09}$ & $\cdots$ & $8.65_{-0.14}^{+0.18}$ & 0.93/123 \\
45 & $6.1_{-1.3}^{+3.7}$ & $0.46_{-0.15}^{+0.09}$ & $\cdots$ & $0.39_{-0.29}^{+0.84}$ & $<0.18$ & $0.28_{-0.11}^{+0.21}$ & $0.20_{-0.17}^{+0.65}$ & $\cdots$ & $0.07_{-0.03}^{+0.08}$ & $\cdots$ & $1.97_{-0.84}^{+2.87}$ & 0.99/54 \\
46 & $18.8_{-2.0}^{+1.9}$ & $1.04_{-0.14}^{+0.40}$ & $\cdots$ & $\cdots$ & $0.34_{-0.34}^{+1.39}$ & $1.09_{-0.30}^{+0.43}$ & $0.93_{-0.21}^{+0.27}$ & $\cdots$ & $\cdots$ & $16.1_{-9.1}^{+15.7}$ & $0.244_{-0.085}^{+0.036}$ & 1.01/132 \\
\end{longtable}

\end{appendix}
\end{document}